\documentclass[preprintnumbers,amsmath,amssymb]{revtex4-1}
\usepackage{mathrsfs}
\usepackage{graphicx}
\usepackage{dcolumn}
\usepackage{bm}
\usepackage{textcomp}
\usepackage{amsmath}
\usepackage[titletoc]{appendix}
\usepackage{color}
\usepackage{multirow}
\usepackage{hyperref}
\hypersetup{                    
	colorlinks,
	citecolor=blue,
	filecolor=blue,
	linkcolor=red,
	urlcolor=blue
}

\newcommand\ket[1]{\ensuremath{|#1\rangle}}

\newcommand\mean[1]{\ensuremath{\langle #1\rangle}}

\newcounter{RomanNumber}

\emergencystretch=\maxdimen
\begin{document}
\title {Universal approach to sending-or-not-sending twin field quantum key distribution}


\author{Xiao-Long Hu,$^{1}$ Cong Jiang,$^{2}$ Zong-Wen Yu,$^{3}$ and Xiang-Bin Wang$^{2,4,5,6,}$}\email{xbwang@mail.tsinghua.edu.cn}

\affiliation{$^{1}$School of Physics, State Key Laboratory of Optoelectronic Materials and Technologies, Sun Yat-sen University, Guangzhou 510275, China\\
	$^{2}$Jinan Institute of Quantum technology, SAICT, Jinan 250101, China\\
	$^{3}$Data Communication Science and Technology Research Institute, Beijing 100191, China\\
	$^{4}$State Key Laboratory of Low Dimensional Quantum Physics, Department of Physics, Tsinghua University, Beijing 100084, China\\
	$^{5}$Synergetic Innovation Center of Quantum Information and Quantum Physics, University of Science and Technology of China, Hefei 230026, China\\
	$^{6}$Shenzhen Institute for Quantum Science and Engineering, and Physics Department, Southern University of Science and Technology, Shenzhen 518055, China}

\begin{abstract}
\textcolor{black}{We present  the method of decoy-state analysis after bit-flip error correction and using confidential observed numbers. Taking this tool} we then construct a universal approach to sending-or-not-sending (SNS) protocol of twin-field quantum key distribution. 
In this improved protocol, the code bits are not limited to heralded events in time windows participated by pulses of intensity $\mu_z$ and vacuum. 
All kinds of heralded events can be used for code bits to distill the final keys.
The number of intensities (3 or 4) and the kinds of heralded events for code bits are automatically chosen by the key rate optimization itself. 
Numerical simulation shows that the key rate rises drastically in typical settings, up to 80\% improvement compared with the prior results. 
Also, larger intensity value can be used for decoy pulses. 
This makes the protocol more robust in practical experiments.
\end{abstract}


\maketitle

\section{Introduction}
Based on principles of quantum mechanics, quantum key distribution (QKD) can provide secure keys for private communication between two parties, Alice and Bob~\cite{bennett1984quantum,lo1999unconditional,shor2000simple,renner2008security,scarani2009security,tomamichel2012tight,xu2020secure,pirandola2020advances}.
In practical implementation, the decoy-state method~\cite{hwang2003quantum,wang2005beating,lo2005decoy,lim2014concise} can be applied for a secure result with imperfect single-photon sources.
And using measurement-device-independent (MDI) QKD protocol~\cite{lo2012measurement,braunstein2012side}, QKD can overcome the security loophole with imperfect detection devices.
Combined with the decoy-state method, MDIQKD can present a secure result with both imperfect single-photon sources and imperfect measurement devices~\cite{tamaki2012phase,wang2013three,curty2014finite,xu2014protocol,yu2015statistical,zhou2016making,hu2021practical,jiang2021higher}.
The decoy-state MDI-QKD protocol has been demonstrated in several experiments~\cite{rubenok2013real,liu2013experimental,yin2016measurement,comandar2016quantum,wang2017measurement,semenenko2020chip,wei2020high,zheng2021heterogeneously}.
So far, applying the 4-intensity protocol~\cite{zhou2016making}, the MDIQKD over a maximum distance of 404 km has been experimentally demonstrated~\cite{yin2016measurement}, while the BB84 QKD has reached a distance record of 421 km~\cite{boaron2018secure} through the decoy-state method.
The channel loss is the major challenge for long-distance QKD given that the key rates of these protocols are limited by the linear bounds of repeaterless QKD, the PLOB bound~\cite{pirandola2017fundamental} established by Pirandola, Laurenza, Ottaviani, and Banchi.
Using a memoryless quantum relay, the twin-field quantum key distribution (TFQKD) proposed recently~\cite{lucamarini2018overcoming} can offer a secure key rate $R$ in the square root scale of channel transmittance $\eta$, i.e., $R\sim O(\sqrt\eta)$.
This makes it possible to greatly improve the performance of QKD at longer distance regimes.
Following this protocol, many variants of TFQKD were proposed~\cite{wang2018twin,tamaki2018information,ma2018phase,cui2019twin,curty2019simple,lu2019improving,maeda2019repeaterless,curras2021tight} and some experiments of TFQKD were demonstrated~\cite{minder2019experimental,liu2019experimental,wang2019beating,zhong2019proof,chen2020sending,liu2021field,chen2021twin,pittaluga2021600,wang2022twin,clivati2022coherent}.
Among those protocols, the sending-or-not-sending (SNS)  protocol~\cite{wang2018twin} has the advantages of MDI security under coherent attacks and it can tolerate large misalignment error. 
So far, the SNS protocol has been extensively studied both theoretically~\cite{yu2019sending,jiang2019unconditional,hu2019sending,xu2020sending,jiang2020zigzag,jiang2021composable,teng2021sending} and experimentally~\cite{minder2019experimental,liu2019experimental,chen2020sending,liu2021field,chen2021twin,pittaluga2021600}.
The method of actively odd parity pairing (AOPP)~\cite{xu2020sending,jiang2020zigzag,jiang2021composable} can further improve the key rate and secure distance of SNS protocol.
Notably, the SNS protocol has been demonstrated in the 511-km field experiment~\cite{chen2021twin} through commercial optical fibers between two metropolitans Jinan and QingDao with MDI security, with Charlie’s measurement station in Mazhan. 
This makes an important proof of the practical applicability of SNS protocol requesting remote single-photon interference with independent lasers.

In previous SNS protocols, only effective events from pulses of intensity $\mu_z$ and vacuum contribute to the final key.
All effective events from decoy pulses are only used for parameter estimation and not used in the key distillation.
Here, 
 we present an improved SNS protocol, in which the code bits are not limited to heralded events in time windows participated by pulses of intensity $\mu_z$ and vacuum. 
All kinds of heralded events can be used for code bits to distill the final key.
\textcolor{black}{The intensities used in code-bit time windows and the numbers of bit values 0 and 1 in code bits cannot be announced in previous protocols, but we need these values in the decoy-state analysis in our improved protocol.
	For this, we will firstly propose an idea of decoy-state analysis after error correction and using confidential observed numbers for our protocol.
	This makes the important preliminary tool for our universal approach to SNS protocol while this itself also makes a general result for the improved decoy-state method.
}
Our improved protocol gives significant rise in the key rate compared with the prior art SNS protocols.
Moreover, in this protocol, larger intensity value can be used for decoy pulses, which makes the protocol more robust in real-world experiments.

This paper is arranged as follows.
\textcolor{black}{In Sec.~\ref{sec:decoy}, we present the method of decoy-state analysis after bit-flip error correction and using the confidential observed numbers.}
In Sec.~\ref{sec:protocol}, we present the improved protocol of SNS TFQKD.
In Sec.~\ref{sec:simulation}, we show the results of numerical simulation of our improved SNS protocol compared with the prior art protocol.
In Sec.~\ref{sec:discussion}, we give discussions about some refined analysis which can further improve the key rate.
The article ends with some concluding remarks in Sec.~\ref{sec:conclusion}.

\textcolor{black}{
\section{Decoy-state analysis after error correction}\label{sec:decoy}
Although there is no way to distill the final key before decoy-state analysis, the bit-flip error correction part alone can be done before decoy-state analysis. 
Here we propose to firstly take the bit-flip error correction and then take the decoy-state analysis. 
After the bit-flip error correction, they (Alice and Bob) can know more observed numbers of specific kinds of bits which are not known to them before error correction and thus makes the decoy-state analysis more effectively. 
With this, they can use all kinds of heralded events for code bits as shown below in our improved protocol. 
Although they observed numbers of all kinds of bits, some of them are confidential observed numbers which can cause information leakage if announced, such as the number of bits with bit value 1. 
If this value is announced, Eve will at least know the parity of all secure bits. 
However, even the confidential values are used or announced in the decoy-study analysis, we can still obtain a secure final key provided that we deduct the final key length by the amount of information leakage for the confidential numbers~\cite{tomamichel2012tight}. 
In particular, we shall use the following} \textcolor{black}{results}:
	\\ \textcolor{black}{{\bf Result 1}: In a QKD protocol with decoy-state method, we can firstly take the bit-flip error correction and then take the decoy-state analysis. 
		In this way, all kinds of observed numbers of bits are known.
\\{\bf Result 2}}: \textcolor{black}{To remove the possible information leakage due to announcement of the number of any kind of bits, we only need to deduct the upper bound amount of information leakage due to that announcement. 
The information leakage $\gamma_\beta$ by announcing a confidential observed number of bits of kind $ \beta$ is bounded by
\begin{equation}
\gamma_\beta \le \log_2 (\overline{m_\beta} - \underline{m_\beta})
\end{equation}
provided that the number $\overline{m_\beta}$ ($\underline{m_\beta}$) upper (lower) bounds the number of bits of kind $\beta$ and $\overline{m_\beta}$ and $\underline{m_\beta}$  can be verified without using any observed confidential numbers. 
For, given that bound values $\overline{m_\beta}$ and $\underline{m_\beta}$, the confidential observed number can always be represented by a $\log_2 (\overline{m_\beta} - \underline{m_\beta})$-length bit string and hence the information leakage in the announcement is upper bounded by 
$\log_2 (\overline{m_\beta} - \underline{m_\beta})$. 
For example, we can upper bound the number of code bits with bit value 1 by $m_t$,  the total number of bits, and lower bound the number of code bits with bit value 1 by 0 without using any confidential observed numbers. If we use several confidential observed numbers in the decoy-state analysis, we have 
\begin{equation}
\Delta \le \sum_\beta \log_2 (\overline{m_\beta} - \underline{m_\beta})
\end{equation} 
to upper bound the total information leakage.
Note that in applying the Result 2, we don’t have to prove whether the announcement of a certain number can indeed cause information leakage. 
In any case we are not sure whether the announcement will cause any information leakage, we can use our Result 2 above to make sure of the security. 
Since all values of $\gamma_\beta$ here are logarithm to a certain natural number, the cost here is negligibly small.
For simplicity, we shall take $\underline{m_\beta}=0$ in the calculation in this work.}

We shall use results above as the preliminary tool to  construct our universal approach to SNS protocol which can use all kinds of heralded events for code bits. 
We emphasize here that the application of our result of decoy-state analysis here is not limited to the improved SNS protocol below, it can in general apply to protocols using vacuum and non-vacuum for bit-value encoding and other kinds of encoding.

\section{The improved protocol of SNS TFQKD}\label{sec:protocol}
The quantum communication part is the same with the existing SNS protocol with AOPP. 
Say, each side uses 4 intensities, $\mu_v=0$, $\mu_x$, $\mu_y$, $\mu_z$, with probabilities $p_v$, $p_x$, $p_y$, $p_z$, respectively. 
They (Alice and Bob) will use those heralded events when Charlie’s measurement device is heralded by one and only one detector for further data processing. 
For ease of presentation, we make the following notations first: \\
Heralded time window: the time window heralded by one and only one detector at Charlie's measurement station, as announced by Charlie; \\
Heralded event: the event produced in a heralded time window; \\
Null time window: the time window when neither of Charlie’s detectors clicks or both of them click;\\
$lr$-event or $lr$-window: an event or a time window when Alice sends out a pulse of intensity $\mu_l$ while Bob sends out a pulse of intensity $\mu_r$;\\
$N_{lr}$: the number of $lr$-windows;\\
$n_{lr}$: the number of heralded $lr$-windows;\\
$N_t$: the total number of time windows in the protocol;\\
$\mean{M}$: the expected value of the quantity $M$.

Encoding: Alice (Bob) regards all heralded windows when she (he) uses intensity $\mu_v$ as a bit value 0 (1) and those when she (he) uses intensity $\mu_l$, with $l\in \{x,y,z\}$, as a bit value 1 (0).
Consequently, a code bit from a heralded time window is a wrong bit when both of them have decided to send out non-vacuum pulses or both of them have decided to send out vacuums.
We define an {\em untagged window} if it's a heralded time window $lv$ or $vr$ with $l, r \in \{x,y,z\}$ and a single photon is actually sent out from users' labs in this time window.
The bits from these untagged windows are defined as {\em untagged bits}.
Note that all untagged bits are right bits.

The main idea of the improved protocol here is that they can use all heralded events to distill the final key.
For this, it differs from the original SNS protocol and the original AOPP-SNS protocol in the secure key length formula, the procedure of classical communication, and the improved decoy-state analysis {\em after error correction}. 
Since our improved protocol is the same with the prior art protocol in the quantum communication part, in what follows we shall focus on the classical communication and data post-processing of our improved protocol. 
First, we consider the improved protocol with original SNS first (Protocol 1) and then we combine the AOPP method (Protocol 1').
We shall then present the decoy-state analysis for our protocol.
As discussed in the appendix, the key rate can be further improved if we take some more refined analysis.

\subsection{Improved protocol with original SNS}	\label{sec:improved_SNS}
Below we shall first present our protocol where bits from all kinds of heralded time windows are regarded as code bits. 
Later, we show that with some modifications, the protocol can also apply to the case where only bits from a specific subset of heralded time windows are regarded as code bits.

1. After the quantum communication, Charlie announces which time windows are heralded windows and which ones are null windows. 
Suppose Charlie has announced $n_t$ heralded time windows and $N_t-n_t$ null time windows. 
They will use those $n_t$ bits from heralded time windows as their code bits. 
They announce the intensities of each one’s pulses sent out in those $N_t-n_t$ null time windows.

2. After error correction to those $n_t$ code bits, Alice knows the positions of all those $n_E$ bit-flip errors and she announces these positions. 
They announce each one’s choice of intensities in the time windows which have produced those $n_E$ bit errors. \\
Remark: Since right bits come from the heralded windows when one and only one of Alice and Bob decides sending, in completion of the steps above, both of them know the positions of time windows of $vv$ and $lr$ with $l,r\in\{x,y,z\}$.
In addition, Alice (Bob) is aware of the positions of time windows of any $lv$ ($vr$). 
This enables them to do the decoy-state analysis.


3. Knowing all time windows of $vv$, $xv$, $yv$, ($vv$, $vx$, $vy$), Alice (Bob) can verify the lower bound of 
$\mean{n_{10}}$ ($\mean{n_{01}}$), the expected value of the number of untagged windows when she (he)  sends out single-photon pulses. 
They publicly announce these bounds.

4. They publicly announce the phase information of all heralded $xx$ windows. 
Those $xx$ windows with the phase slice $(\theta_A - \theta_B)$ of states $\ket{\mu_x e^{i \theta_A}} \ket{\mu_x e^{i \theta_B}}$ satisfying the condition 
\begin{equation}\label{equ:postselection}
	1 - |\cos (\theta_{A} - \theta_{B})| \le \lambda
\end{equation}
will be used to verify the lower bound of $e_1^{ph}$, the phase-flip error rate of untagged bits in the decoy-state analysis.
Here, $\lambda$ is a positive number close to 0 and its value is determined by Alice and Bob according to the result of channel test and calibration in the experiment to obtain a satisfactory key rate.

5. They calculate the final key length for SNS protocol by 
\begin{equation}\label{equ:tilde_n}
	\tilde{n} = n - \Delta 
\end{equation}
with
\begin{equation}\label{equ:keylength}
	n= n_1 [1 - H(e_1^{ph})] - f n_t H(E_t)  - 2\left(\log_2\frac{2}{\varepsilon_\text{cor}} - 2 \log_2\frac{1}{\sqrt{2}\varepsilon_\text{PA}\hat\varepsilon}\right),
\end{equation}
where $n_1$ is the number of untagged bits from those code bits, and $E_t$ is the quantum bit-flip error rate (QBER) of all code bits before distillation.
Bounds of $n_1$ and $e_1^{ph}$ can be verified by decoy-state analysis shown in subsection~\ref{subsec:decoy}, and the values of $n_t$ and $E_t$ can be directly observed in the experiment.
And $f$ is the error correction inefficiency, $H(x)=-x\log_2(x)-(1-x)\log_2(1-x)$ is the Shannon entropy, $\varepsilon_\text{cor}$ is the failure probability of error correction, $\varepsilon_\text{PA}$ is the failure probability of privacy amplification, $\hat\varepsilon$ is the coefficient while using the chain rules of max- and min-entropy~\cite{jiang2019unconditional}, and as shall be studied in detail, $\Delta$ is the additional information leakage of the final key due to classical communication in Step 3 above. 
\textcolor{black}{The value is quite small and can be upper bounded by our Result 2 in Sec.~\ref{sec:decoy}.}
	 According to Ref.~\cite{tomamichel2012tight}, in obtaining the secure final key, one has to remove all information leakage to private raw bits in classical communication. 
Here, in our protocol, besides the classical information for error correction, the classical communication in Step 3 can cause information leakage of the private raw bits, denoted by $\Delta$.
\textcolor{black}{As shown in Sec.~\ref{sec:decoy},}
a loose upper bound of $\Delta$ takes the magnitude order of $\log_2 (\sum_l n_{lv} +\sum_r n _{vr})$.
Straightly, if we disregard $\Delta$, the security of Eq.~\eqref{equ:keylength} can be shown in a similar way applied in the original SNS protocol~\cite{wang2018twin,jiang2019unconditional}.

In doing the error test for the bit-flip error correction in Step 2, they have to randomly take a small fraction $\delta$ of time windows to test the QBER of their code bits. 
They have to discard the events of these time windows. 
Since they can verify faithfully the fact of zero error after error correction, we shall always simply take the small value $\delta=0$ in our calculation. 
In the error correction, we assume Alice to be the party that computes the positions of those $n_E$ bit-flip errors.

Different from the existing SNS protocol, here they can count in all heralded events for code bits to distill the final key.
Surely, they can also choose to only use part of heralded events, e.g., limiting $l,r$ in $\{x,y\}$, $\{y,z\}$, or $\{z\}$ only, for code bits.
For an advantageous final key rate, we can take different options in choosing different subsets of heralded events for code bits under different conditions. 
We use notation $[x,y,z]$ for the code-bit option that they use all heralded events as their code bits. 
In such an option, 
mathematically we have 
\begin{equation}\label{equ:ntEt_xyz}
		n_t = \sum_{l,r\in \{x,y,z,v\}} n_{lr},	\ E_t = \left(n_{vv} + \sum_{l,r\in \{x,y,z\}} n_{lr}\right) / n_t.
\end{equation} 
We use notation $[y,z]$ for the code-bit option that they limit the code bits to those bits from heralded ${lr}$-windows with $l,r \in \{y,z,v\}$ only.  
In such an option, mathematically we have 
\begin{equation}\label{equ:ntEt_yz}
		n_t = \sum_{l,r\in \{y,z,v\}} n_{lr},	\	E_t = \left(n_{vv} + \sum_{l,r\in \{y,z\}} n_{lr}\right) / n_t.
\end{equation} 
Also, we use notation $[z]$ for the code-bit option that they limit the code bits to those bits from heralded ${lr}$-windows with $l,r \in \{z,v\}$ only.
In such an option, mathematically we have
\begin{equation}\label{equ:ntEt_z}
	n_t = \sum_{l,r\in \{z,v\}} n_{lr}, \ E_t = (n_{vv} + n_{zz}) / n_t.
\end{equation}
If they take the option $[y,z]$ ($[z]$), in addition to those contents in Step 1 of Protocol 1, they each needs to announce all time windows when she (he) has sent out pulses of intensity $\mu_x$ ($\mu_x$ or $\mu_y$) so that they become aware of which bits are code bits in the option $[y,z]$ ($[z]$). 
In the subsequent calculations, the relevant quantities such as $n_1$ and $e_1^{ph}$, are now redefined based on the code-bit option $[y,z]$ ($[z]$).
Note that $n_t$ and $E_t$ can be directly observed in the experiment.
Formulas in Eqs.~\eqref{equ:ntEt_xyz}~\eqref{equ:ntEt_yz}~\eqref{equ:ntEt_z} are used to explain which heralded events are contained in $n_t$ and $E_t$ and these formulas can be used to calculate the expected observed values in the numerical simulation.

\subsection{Improved protocol with AOPP-SNS}	\label{sec:improved_AOPP}
Surely, we can apply AOPP~\cite{xu2020sending,jiang2020zigzag,jiang2021composable} to our method above for a better performance of the whole protocol. 
Alice makes random odd-parity bit pairs from her code bits, Bob takes parity check to all pairs and then they take one bit randomly in any pair which has passed the parity check. 
We can expect a much lower QBER $E^\prime_t$ in those survived bits after AOPP.

Compared with Protocol 1 for the original SNS, there are two major differences here: 
First, to reduce the bit-flip error rate, they have to take  the subprotocol of post-selection with AOPP, denoted as subprotocol $\mathcal A$. 
Second, to calculate the final key, they need to calculate the number of untagged bits in those post-selected $n_t^\prime$ code bits after subprotocol $\mathcal A$.

\subsubsection{Subprotocol $\mathcal A$: post-selection with AOPP}\label{subsubsec:subprotocol A}
Take the option of $[x,y,z]$ as an example, in the AOPP, Alice first makes $n_{\text{odd}}=\min\{n_x^A+n_y^A+n_z^A, n_v^A\}$ odd-parity pairs, where $n_l^A$ is the number of effective time windows when Alice uses intensity $\mu_l$ with $l=v,x,y,z$. 
Specifically, if $n_x^A+n_y^A+n_z^A \ge n_v^A$, she randomly chooses $n_{\text{odd}}=n_v^A$ bits from those $n_x^A+n_y^A+n_z^A$ bits with bit value 1 and then makes $n_{\text{odd}}$ random odd-parity pairs; if $n_x^A+n_y^A+n_z^A < n_v^A$, she randomly chooses $n_{\text{odd}}=n_x^A+n_y^A+n_z^A$ bits from those $n_v^A$ bits with bit value 0 and then makes $n_{\text{odd}}$ random odd-parity pairs.
She announces the positions of each pair.
Among these $n_{\text{odd}}$ pairs, $n_t^\prime$ of them have odd parity at Bob’s side and these $n_t^\prime$ pairs will pass the parity check by Bob. 
Surely, only two kinds of pairs can pass the parity check: a pair containing two bit-flip errors or a pair containing no bit-flip error. 
For ease of presentation, we call it a {\em right pair} if there is no bit-flip error in that pair.
The value of $n_t^\prime$ is an experimentally observed value to Alice and she needs classical communication with Bob in doing the parity check.
Then,  they take one bit randomly in each pair which has passed the parity check.
A right pair will produce a right bit.

Here, Alice takes error correction to those $n_t^\prime$ survived code bits after AOPP. 
She computes the positions of wrong bits and publicly announces them.
This means that they become aware of all those wrong bits in those survived pairs after the parity check of AOPP. 
The numbers of wrong bits and right bits are $n_E^\prime$ and $n_R^\prime=n_t^\prime-n_E^\prime$, respectively.

\subsubsection {Key length calculation}
Given announced information above, both of Alice and Bob know the values of $n_{vv}$ and $n_{lr}$ with $l,r\in\{x,y,z\}$, Alice knows the values of ${n_{xv}}$, ${n_{yv}}$, and ${n_{zv}}$, and Bob knows the values of ${n_{vx}}$, ${n_{vy}}$, and ${n_{vz}}$. 
Alice (Bob) can calculate the non-asymptotic lower bound of $\mean{n_{10}}$ ($\mean{n_{01}}$) by decoy-state analysis and announce it.
Then, they can lower bound the value $n_1^\prime$, the number of survived untagged bits after AOPP, by method in Ref.~\cite{jiang2021composable} using the total number of code bits and untagged code bits before AOPP.  
In addition, they know all those heralded events of $xx$-windows which satisfy the phase slice condition in Eq.~\eqref{equ:postselection}. 
With these,  they can upper bound the non-asymptotic  value of $e_1^{ph}$ by decoy-state analysis and also $e_1^{\prime ph}$, the phase-flip error rate of survived code bits after AOPP, with iteration formulas in Ref.~\cite{jiang2021composable}.
Since the values of $\mean{n_{10}}$ and $\mean{n_{01}}$ are announced, there is information leakage $\Delta$ and the final key length is
\begin{equation}\label{equ:tilde_n_AOPP}
	\tilde{n}^\prime = n^\prime -\Delta
\end{equation}
with
\begin{equation}\label{equ:keylength_AOPP}
	n^\prime= n_1^\prime [1 - H(e_1^{\prime ph})] - f n_t^\prime H(E_t^\prime)  - 2\left(\log_2\frac{2}{\varepsilon_\text{cor}} - 2 \log_2\frac{1}{\sqrt{2}\varepsilon_\text{PA}\hat\varepsilon}\right).
\end{equation}
Here, $n_t^\prime$ is the number of survived code bits after AOPP and $E^\prime_t$ is the QBER in those survived code bits after AOPP as introduced above.
The values of them can be directly observed after the parity check of AOPP.
As usual, in our calculation, we omit the small fraction of bits cost in testing the QBER.

\subsubsection {Protocol 1'}
For completeness, we write the following improved protocol of AOPP-SNS in the code-bit option $[x,y,z]$, naming as Protocol 1’:

1. Same as Step 1 in Protocol 1.   

2. They take subprotocol $\mathcal A$ to post-select $n_t^\prime$ code bits whose QBER $E^\prime_t$ is supposed to be significantly lower than that before this post-selection. 

3. After error correction to those $n_t^\prime$ post-selected code bits, Alice knows the positions of all those $n_E^\prime$ bit-flip errors and she announces these positions. 
They announce each one’s intensities of pulses in all heralded time windows except those producing $2n_R^\prime$ code bits in $n_R^\prime$ right pairs.
After this, both of them know the positions of time windows of $vv$ and $lr$ with $l,r\in\{x,y,z\}$, since all bits of right pairs come from heralded windows when one and only one of Alice and Bob decides sending.
In addition, Alice (Bob) is aware of the positions of time windows of any $lv$ ($vr$). 

4. Same as Step 3 in Protocol 1.  

5. Same as Step 4 in Protocol 1.  

6. They calculate the key length by Eqs.~\eqref{equ:tilde_n_AOPP}~\eqref{equ:keylength_AOPP}.

Protocol 1' can be modified for code-bit options $[y,z]$ and $[z]$.
Since they shall take further processing to code bits in option $[y,z]$ (or $[z]$), besides the contents in Step 1 of Protocol 1’, they each needs to announce in which heralded time windows she/he has chosen pulse intensity $\mu_x$ in option $[y,z]$ ($\mu_x$ or $\mu_y$ in option $[z]$).  
With this, they know which bits are code bits in their code-bit option. Also, in the subsequent calculations, the relevant quantities such as $n_t$, $n_t^\prime$, $E_t$, $E_t^\prime$, $n_1$, $n_1^\prime$, $e_1^{ph}$, and $e_1^{\prime ph}$ are now defined based on code bits in option $[y,z]$ or $[z]$, respectively.

\subsection{Decoy-state analysis and $\Delta$ term in the key length formula}	\label{subsec:decoy}
Here the mathematical formulas are the same with the existing ones:
\begin{equation}\label{equ:s10L}
	\mean{s_{10}} \ge \mean{s_{10}}^L = \frac{e^{\mu_x} \mu_y^2 \mean{S_{xv}} - e^{\mu_y} \mu_x^2 \mean{S_{yv}} - (\mu_y^2-\mu_x^2) \mean{S_{vv}}}{\mu_x\mu_y(\mu_y-\mu_x)},
\end{equation}
\begin{equation}\label{equ:s01L}
	\mean{s_{01}} \ge \mean{s_{01}}^L = \frac{e^{\mu_x} \mu_y^2 \mean{S_{vx}} - e^{\mu_y} \mu_x^2 \mean{S_{vy}} - (\mu_y^2-\mu_x^2) \mean{S_{vv}}}{\mu_x\mu_y(\mu_y-\mu_x)},
\end{equation}
and
\begin{equation}\label{equ:eph}
	\mean{e_1^{ph}} \le \mean{e_1^{ph}}^U = \frac{\mean{T_X} - e^{-2\mu_x} \mean{S_{vv}}/2}{2\mu_x e^{-2\mu_x} \mean{s_1}},
\end{equation}
where $\mean{s_{10}}$ ($\mean{s_{01}}$) are the expected value of counting rate of time windows when Alice (Bob) sends out a single-photon pulse and Bob (Alice) sends out vacuum, $\mean{S_{lr}}$ is the expected value of counting rate of $lr$-windows, $\mean{s_{1}}=(\mean{s_{10}}+\mean{s_{01}})/2$ is the expected value of counting rate of all single-photon events, and $\mean{T_X}$ is the expected value of error counting rate of $xx$ windows which satisfy the phase slice condition in Eq.~\eqref{equ:postselection}.
However, since they do the analysis with classical communications above, Alice (Bob) knows the observed values of $n_{vv}$ and $n_{lv}$ for $l\in \{x, y, z\}$ ($n_{vr}$ for $r\in \{x, y, z\}$) from {\em all} time windows. 
She (He) can directly use these as the input values in Eqs.~\eqref{equ:s10L}~\eqref{equ:s01L}~\eqref{equ:eph} above, say, $S_{vv}= n_{vv}/N_{vv}$ and $S_{lv}= n_{lv}/N_{lv}$ ($S_{vr}= n_{vr}/N_{vr}$).
Thus, in doing the decoy-state analysis after error correction, they don't have to reserve some time windows of $vv$, $lv$, and $vr$ as random samples to test the observed values of $S_{vv}$, $S_{lv}$, and $S_{vr}$.
Then, the bound of $\mean{S_{lr}}$ can be calculated from the observed value ${S_{lr}}$ by Chernoff bound~\cite{chernoff1952measure} in the Appendix~\ref{sec:chernoff}.
With the lower bound of $\mean{s_{10}}$, Alice can calculate the bound of $\mean{n_{10}}$ by:
\begin{equation}\label{equ:n10}
	\mean{n_{10}} = \sum_l (N_{lv} e^{-\mu_l}\mu_l) \mean{s_{10}},
\end{equation}
where the summation of $l$ depends on the chosen option, i.e., $l\in \{x, y, z\}$ with the option $[x,y,z]$, $l\in \{y, z\}$ with the option $[y,z]$, or $l\in \{z\}$ with the option $[z]$.
Similarly, Bob can obtain the bound of $\mean{n_{01}}$:
\begin{equation}\label{equ:n01}
	\mean{n_{01}} = \sum_r (N_{vr} e^{-\mu_r}\mu_r) \mean{s_{01}}.
\end{equation}
After announcing the bounds of $\mean{n_{10}}$ and $\mean{n_{01}}$, they can calculate $\mean{n_{1}}=\mean{n_{10}}+\mean{n_{01}}$ and then obtain $n_1$ by using Chernoff bound again.

In calculating the lower bound of $\mean{n_{10}}$, Alice has used the values of ${n_{xv}}$, ${n_{yv}}$, and ${n_{vv}}$. 
These values are related to the number of raw bits with bit values 1. 
The exact number of bit value 1 can make extra information leakage because there is no way to know this in advance for anyone in an entanglement purification protocol. 
In Step 3 above in the classical communication, though Alice does not announce the values of ${n_{xv}}$, ${n_{yv}}$, and ${n_{vv}}$, she has to announce $\underline{\mean{n_{10}}}$, her calculated lower bound of $\mean{n_{10}}$, which is dependent on the values  of ${n_{xv}}$, ${n_{yv}}$, and ${n_{vv}}$, which causes information leakage of those raw bits for final key distillation.
However, we can upper bound the  amount of information leakage by using $\mean{n_{10}}^u$, the upper bound of $\mean{n_{10}}$, known to Eve even if Alice does not announce anything in this step. 
Say, Eve had a prior information that $0\le \underline{\mean{n_{10}}} \le \mean{n_{10}}^u$ before Alice's announcement. 
This means that Alice’s announcement of $\underline{\mean{n_{10}}}$ can be represented by a bit string  not longer than $\log_2 \mean{n_{10}}^u$ bits, and hence the information leakage of the untagged bits is not larger than $\log_2 \mean{n_{10}}^u$ bits. 
Similarly we can also bound the information leakage due to Bob's announcement in this step by introducing  $\mean{n_{01}}^u$, the upper bound of $\mean{n_{01}}$, known to Eve without the Bob’s announcement.
Therefore, we have:
\begin{equation}\label{equ:Delta}
	\Delta \le \log_2 \mean{n_{10}}^u + \log_2 \mean{n_{01}}^u.
\end{equation}
Based on this, we can simply choose the following loose bound $\Delta \le 2 \log_2 (n_t - n_{vv} - \sum_{l^\prime, r^\prime} n_{l^\prime r^\prime})$ where $n_t$ is the total number of heralded time windows and $l^\prime,r^\prime \in \{x,y,z\}$. 
Though there are obviously tighter bounds for $\Delta$,  such a loose bound is quite good already given its logarithm form.

Remark 1: We don’t have to worry about the information leakage of private raw bits due to the announcement in Steps 2 there. 
Those announcement are only related to tagged bits only, instead of private raw bits. 
Note that in our key length formula, we have assumed all tagged bits are known to Eve, and thus there is no extra information leakage in this process. 

Remark 2: If we want to use the joint constraints of statistical fluctuation~\cite{yu2015statistical} in the decoy-state analysis of calculating $\mean{s_{1}}=(\mean{s_{10}}+\mean{s_{01}})/2$, 
the values of $n_{vx}$, $n_{vy}$, $n_{xv}$ and $n_{yv}$ are required to be announced.
If we use $[x,y,z]$ for code bits, this announcement introduces the extra information leakage 
\begin{equation}\label{equ:Delta_jc}
	\Delta_{\text{jc}}=\log (n_{vx}^u) + \log (n_{xv}^u) + \log (n_{vy}^u) + \log (n_{yv}^u)\le 4 \log_2 (n_t - n_{vv} - \sum_{l^\prime, r^\prime \in \{x,y,z\}} n_{l^\prime r^\prime}),
\end{equation}
where $n_{lr}^u$ is a upper bound of $n_{lr}$ known to Eve.
And if $[y,z]$ is used, the extra information leakage introduced is \begin{equation}\label{key}
	\Delta_{\text{jc}}=\log (n_{vy}) + \log (n_{yv})\le 2 \log_2 (n_t - n_{vv} - \sum_{l^\prime, r^\prime \in \{y,z\}} n_{l^\prime r^\prime}).
\end{equation}
In this case, the final key length is $\tilde{n}  = n - \Delta_{\text{jc}}$.

\section{Numerical Simulation}\label{sec:simulation}

In this part, we show the numerical results of our improved AOPP-SNS protocol, and compare them with the results of the prior art AOPP-SNS protocol~\cite{jiang2020zigzag,jiang2021composable}.
The results will be shown in the form of key rate per pulse, i.e. $R=\tilde{n}^\prime/N_t$.
The device parameters used in the simulation are listed in Table~\ref{tab:parameters}.
We shall estimate what values would be probably observed in the normal cases by the linear models as previously.
At each distance, the optimization is taken over the values of $p_x$, $p_y$, $p_z$ and $\mu_x$, $\mu_y$, $\mu_z$ and choosing heralded events $[x,y,z]$ or $[y,z]$ for raw bits by the advantageous key rate.
The optimization can set $p_z=0$ at some distances and the protocol automatically becomes a 3-intensity protocol.
\begin{table}[htbp]
	\centering
	\begin{tabular}{cccccc}
		\hline
		$d$         &   $e_d$   &   $\eta_d$    &   $f$   &    $\xi$  &  $\alpha$ \\
		\hline
		$10^{-9}$  &   1.5\%     &   50\%        &   1.1     &   $10^{-10}$      &  0.2dB/km\\
		\hline
	\end{tabular}
	\caption{\label{tab:parameters}
		Devices' parameters used in numerical simulations.
		$d$ is the dark count rate per pulse of each detector at Charlie's side;
		$e_d$ is the misalignment error in $X$ windows;
		$\eta_d$ is the detection efficiency of each detector at Charlie's side;
		$f$ is the error correction inefficiency;
		$\xi$ is the failure probability in the parameter estimation;
		$\alpha$ is the channel loss.}
\end{table}

\begin{figure}[htbp]
	\centering\includegraphics[width=250pt]{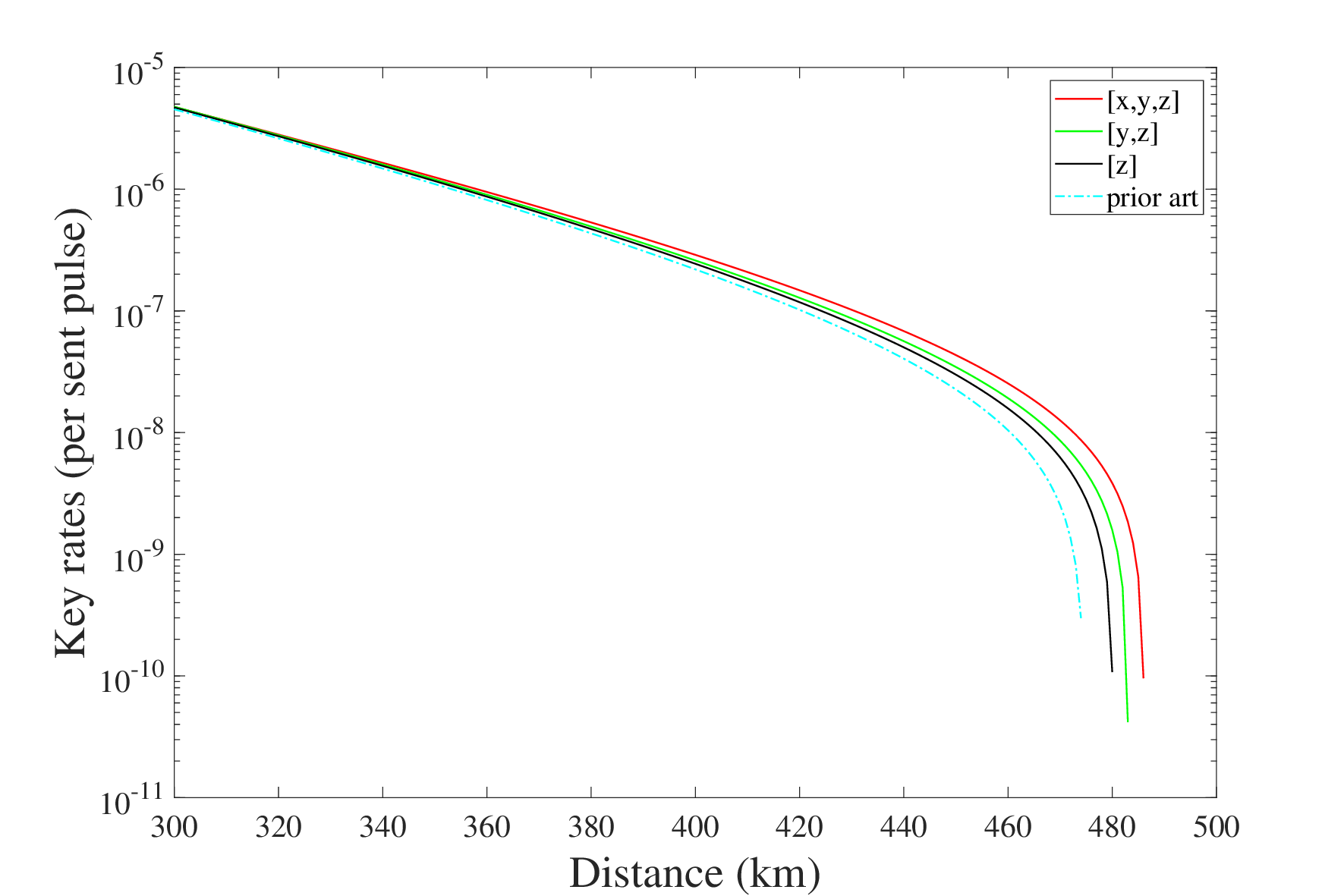}
	\caption{The optimized key rates (per pulse pair) versus transmission distance with different heralded events counted in $n_1^\prime$. Here, we set $N_t=10^{11}$. 	``Prior art'': key rate of the SNS~\cite{wang2018twin} protocol with AOPP~\cite{xu2020sending} using the existing 4-intensity protocol~\cite{yu2019sending,jiang2019unconditional} which has been applied in the 509 km experiment~\cite{chen2020sending} and 511 km field test between metropolitans~\cite{chen2021twin}. In all our calculations, the strict finite-key effects are taken into consideration by applying the method in Ref.~\cite{jiang2020zigzag,jiang2021composable}.}\label{fig:simulation1}
\end{figure}

We show the optimized key rates versus transmission distance in Fig.~\ref{fig:simulation1} and Tables~\ref{tab:simulation1}~\ref{tab:simulation2}.
\begin{table}[htbp]
	\centering
		\begin{tabular}{ccccc}
			\hline
																				& $200$ km						&   $300$ km 					&  $400$ km						&	$450$km \\
			\hline
			$[x,y,z]$ 												&  $5.99\times10^{-5}$ & $4.74\times10^{-6}$ & $2.89\times10^{-7}$	 & $4.33\times10^{-8}$\\
			$[y,z]$													&  $6.14\times10^{-5}$ & $4.76\times10^{-6}$ & $2.59\times10^{-7}$	& $3.46\times10^{-8}$\\
			$[z]$ 													&  $6.11\times10^{-5}$ & $4.69\times10^{-6}$ & $2.44\times10^{-7}$	& $2.99\times10^{-8}$\\
			prior art 	&  $6.03\times10^{-5}$ & $4.53\times10^{-6}$ & $2.19\times10^{-7}$	& $2.26\times10^{-8}$\\
			\hline
		\end{tabular}
		\caption{\label{tab:simulation1}The optimized key rates (per pulse pair) at some transmission distance with different heralded events counted in $n_1^\prime$. Here, we set $N_t=10^{11}$.}
\end{table}
\begin{table}[htbp]
	\centering
	\begin{tabular}{ccccc}
		\hline
		& $200$ km						&   $300$ km 					&  $350$ km						&	$380$km \\
		\hline
		$[x,y,z]$ 												&  $4.76\times10^{-5}$ & $2.95\times10^{-6}$ & $4.83\times10^{-7}$	 & $8.12\times10^{-8}$\\
		$[y,z]$													&  $4.78\times10^{-5}$ & $2.64\times10^{-6}$ & $3.83\times10^{-7}$	& $4.56\times10^{-8}$\\
		$[z]$ 													&  $4.71\times10^{-5}$ & $2.49\times10^{-6}$ & $3.34\times10^{-7}$	& $2.88\times10^{-8}$\\
		prior art 	&  $4.55\times10^{-6}$ & $2.25\times10^{-6}$ & $2.64\times10^{-7}$	& $4.86\times10^{-9}$\\
		\hline
	\end{tabular}
	\caption{\label{tab:simulation2}The optimized key rates (per pulse pair) at some transmission distance with different heralded events counted in $n_1^\prime$. Here, we set $N_t=10^{10}$.}
\end{table}
From these results, 
we can find that the key rates in the option $[y,z]$  are always higher than those in $[z]$ or prior art non-asymptotic results
of the SNS~\cite{wang2018twin} protocol with AOPP~\cite{xu2020sending} using the existing 4-intensity protocol~\cite{yu2019sending,jiang2019unconditional} which has been applied in the 509 km experiment~\cite{chen2020sending} and 511 km field test between metropolitans~\cite{chen2021twin} (labeled as ``prior art'' in the figures and tables). 
In all our calculations, the strict finite-key effects are taken into consideration by applying the method in Ref.~\cite{jiang2020zigzag,jiang2021composable}.
Both options of  $[x,y,z]$ and $[y,z]$ can present advantageous results at different distances.
Especially when the total number of pulse pairs is small and the communication distance is long, our improved SNS protocol works much better than the prior protocol, e.g. 83\% higher at the distance of 350 km with $N_t=10^{10}$.
This improvement makes the SNS protocol more practical for the real-life quantum communication, since the communication time is usually short and the total number of pulse pairs is usually small.
At a certain distance with given $N_t$, we can choose either $[y,z]$ or $[x,y,z]$ for key distillation, depending on an advantageous key length result.

\begin{figure}[htbp]
	\centering\includegraphics[width=250pt]{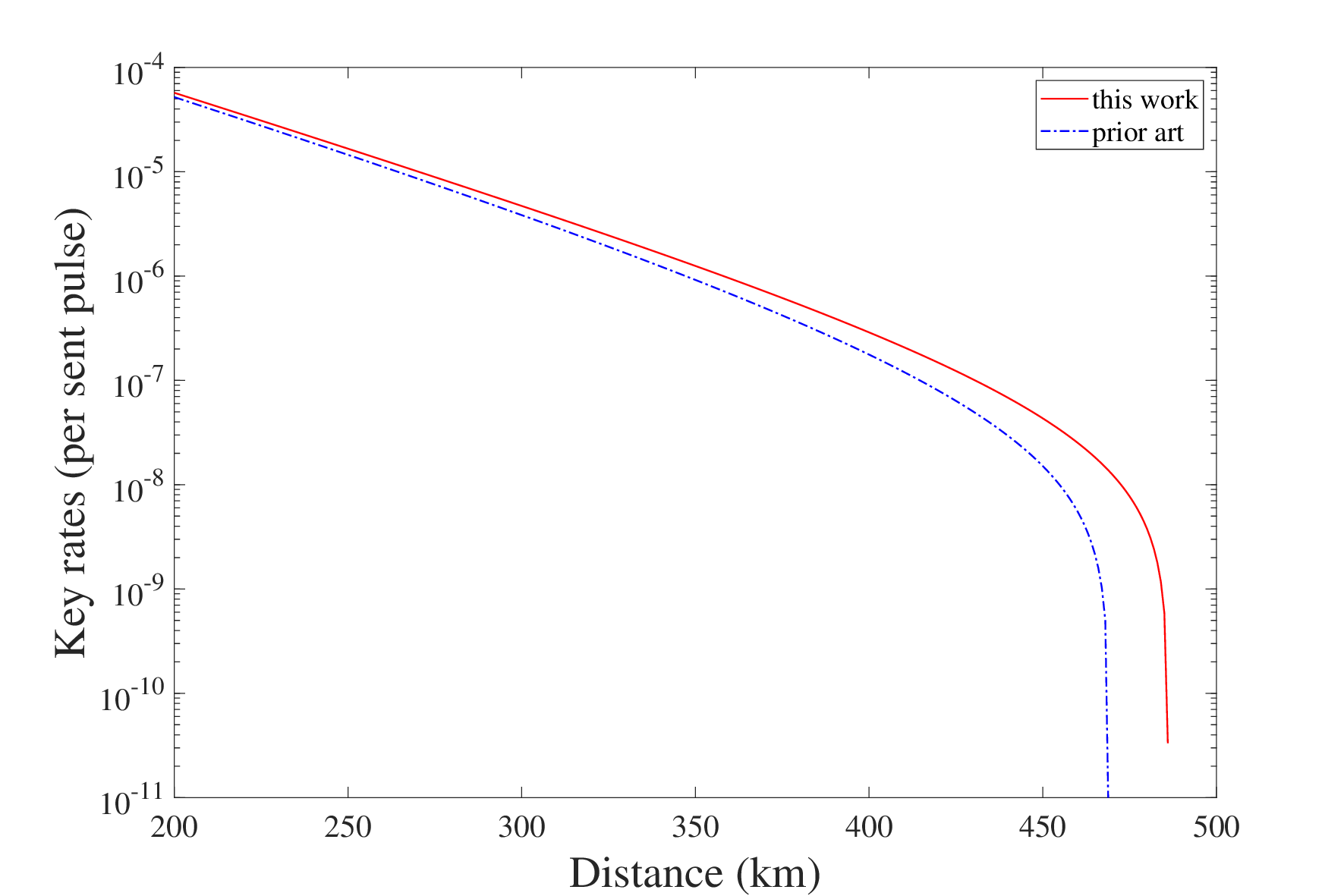}
	\caption{The optimized key rates (per pulse pair) versus transmission distance with different protocols. Here, we fix $\mu_x=0.2$ and set $N_t=10^{11}$.}\label{fig:fixmux}
\end{figure}

Moreover, we show the numerical results with fixed intensity $\mu_x=0.2$ through different protocols in Fig.~\ref{fig:fixmux} and Table~\ref{tab:fixmux}.
Decoy pulses with such large intensity is easier to prepare in real-life experiments, compared with the optimal intensity less than 0.1 used in previous protocols.
In this case, the key rate of the option $[x,y,z]$ is always higher than that of $[y,z]$.
Thus, we choose the option $[x,y,z]$ in this simulation.
When fixing $\mu_x=0.2$, our improved protocol works much better than the prior one.
\begin{table}[htbp]
	\centering
	\begin{tabular}{ccccc}
		\hline
		& $200$ km						&   $300$ km 					&  $400$ km						&	$450$km \\
		\hline
		this work 												&  $5.72\times10^{-5}$ & $4.71\times10^{-6}$ & $2.89\times10^{-7}$	 & $4.33\times10^{-8}$\\
		prior art 	&  $5.20\times10^{-5}$ & $3.84\times10^{-6}$ & $1.77\times10^{-7}$	& $1.51\times10^{-8}$\\
		\hline
	\end{tabular}
	\caption{\label{tab:fixmux}The optimized key rates (per pulse pair) at some transmission distance with different protocols. Here, we fix $\mu_x=0.2$ and set $N_t=10^{11}$.}
\end{table}

\begin{figure}[htbp]
	\centering\includegraphics[width=250pt]{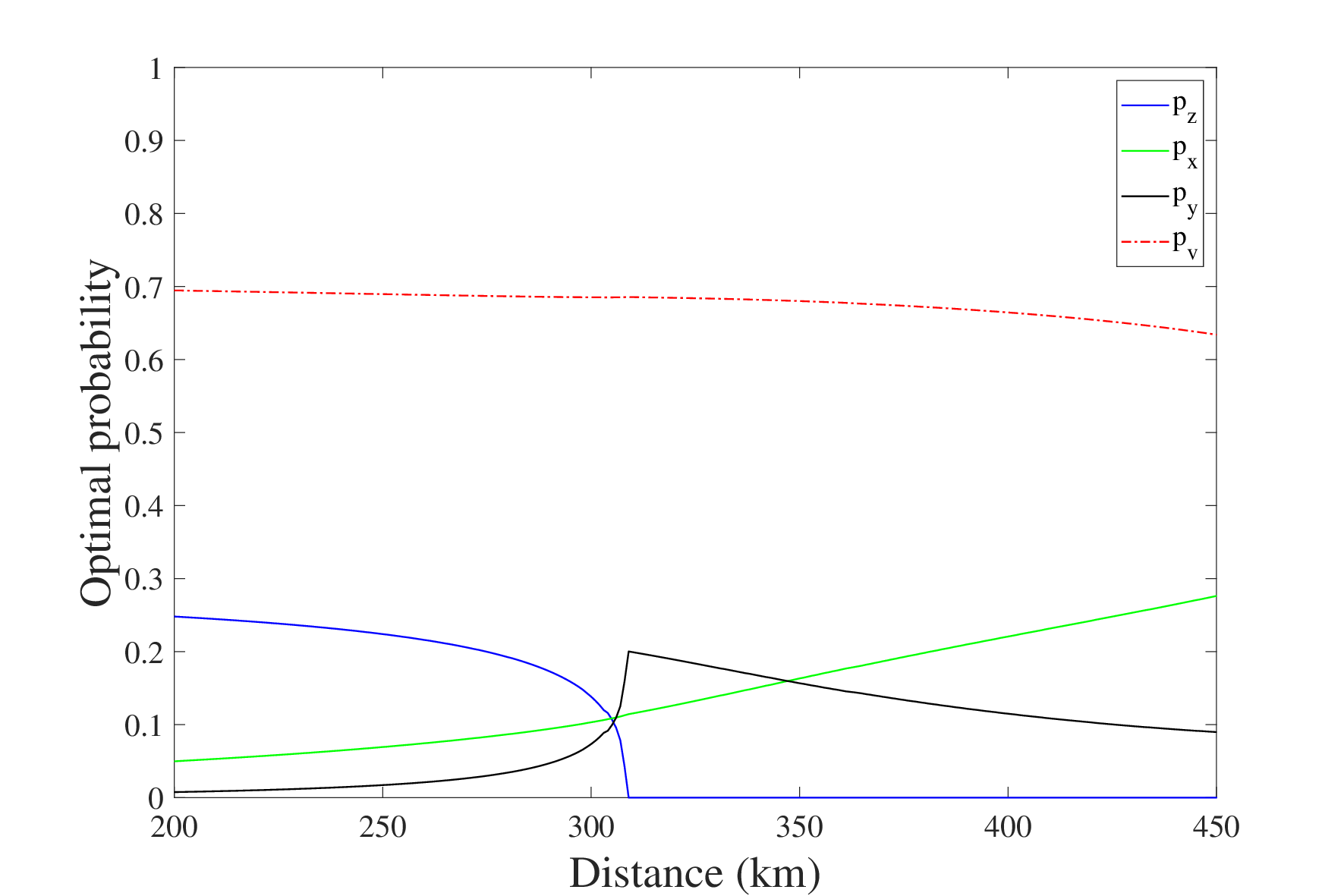}
	\caption{The optimal probabilities for different intensities versus transmission distance. Here, we set $N_t=10^{11}.$}\label{fig:pro}
\end{figure}

In Fig.~\ref{fig:pro}, we show the optimal probabilities versus transmission distance.
When the distance goes large, the optimal probability for intensity $\mu_z$ goes to 0, and our protocol becomes 3-intensity protocol automatically.
But this 3-intensity protocol is different from the existing 3-intensity protocol~\cite{yu2019sending,liu2021field}: in this protocol, we shall use the sending of both no-zero intensities ($\mu_x$ and $\mu_y$) of pulses for code bits, while the prior 3-intensity protocol only uses the sending of one intensity ($\mu_y$) for code bits and we use all time windows instead of reserving some time windows for test. 
With higher probabilities for intensities $\mu_x$ and $\mu_y$, the effect of the statistical fluctuation is still small even when the total number is small and the communication distance is long.
At the same time, the heralded events from sources of intensities $\mu_x$ and $\mu_y$ can used for key distillation.
This helps our protocol work well in practical scenarios with few pulses and long distances.

As shown in Appendix~\ref{sec:discussion}, the key rate can be further improved if we take more refined analysis.

\section{Conclusion}\label{sec:conclusion}
In this paper, we proposed an improved SNS protocol, in which the code bits are not limited to heralded events in time windows participated by pulses of intensity $\mu_z$ and vacuum. 
All kinds of heralded events can be used for code bits to distill the final key.
Our protocol performs well even when the total number of pulse pairs is small and the intensity of decoy pulses is large.
This makes our protocol more practical and robust in real-life quantum communication.

\section*{Acknowledgments}
We acknowledge the financial support in part by the Ministration of Science and Technology of China through the National Key Research and Development Program of China Grant No. 2017YFA0303901 and National Natural Science Foundation of China Grants No. 12147107, No. 11774198, and No. 11974204.

\appendix
\section{Chernoff bound}\label{sec:chernoff}
\textcolor{black}
{
	We can use the Chernoff bound to estimate the expected value with their observed values~\cite{chernoff1952measure}. 
	We denote $X_1,X_2,\dots,X_n$ as $n$ random samples, whose values are 1 or 0, and $X$ as their sum satisfying $X=\sum_{i=1}^nX_i$. 
	We denote $E$ as the expected value of $X$.}
We have
\begin{align}
	\label{EL}E^L(X,\xi)=&\frac{X}{1+\delta_1(X,\xi)},\\
	\label{EU}E^U(X,\xi)=&\frac{X}{1-\delta_2(X,\xi)},
\end{align}
where $\delta_1(X,\xi)$ and $\delta_2(X,\xi)$ are the positive solutions of the following equations:
\begin{align}
	\label{delta1}\left(\frac{e^{\delta_1}}{(1+\delta_1)^{1+\delta_1}}\right)^{\frac{X}{1+\delta_1}}&=\xi,\\
	\label{delta2}\left(\frac{e^{-\delta_2}}{(1-\delta_2)^{1-\delta_2}}\right)^{\frac{X}{1-\delta_2}}&=\xi,
\end{align}
where $\xi$ is the failure probability.

Besides,  the Chernoff bound can be used to estimate their real values with their expected values. 
Similar to Eqs.~\eqref{EL}-\eqref{delta2}, the real value, $O$, can be estimated by its expected value, $Y$:
\begin{align}
	\label{OU}&O^U(Y,\xi)=[1+\delta_1^\prime(Y,\xi)]Y,\\
	\label{OL}&O^L(Y,\xi)=[1-\delta_2^\prime(Y,\xi)]Y,
\end{align}
where $\delta_1^\prime(Y,\xi)$ and $\delta_2^\prime(Y,\xi)$ are the positive solutions of the following equations:
\begin{align}
	\left(\frac{e^{\delta_1^\prime}}{(1+\delta_1^\prime)^{1+\delta_1^\prime}}\right)^{Y}&=\xi,\\
	\label{endd}\left(\frac{e^{-\delta_2^\prime}}{(1-\delta_2^\prime)^{1-\delta_2^\prime}}\right)^{Y}&=\xi.
\end{align}

\section{Some more refined analysis}\label{sec:discussion}


\subsection{Refined QBER} 
When the option of $[x,y,z]$ or $[y,z]$ is used, Alice can observe QBERs of different kinds of code bits. 
Take the option of  $[x,y,z]$ for an example, if we use notations $E_v^A$, $E_x^A$, $E_y^A$, and $E_z^A$ for her observed QBERs of code bits in heralded time windows when she sends out pulses of intensities $\mu_v$, $\mu_x$, $\mu_y$,  and $\mu_z$, respectively, and notations $n_v^A$, $n_x^A$, $n_y^A$, $n_z^A$ for the numbers of these 4 kinds of code bits, we can replace Eq.~\eqref{equ:ntEt_xyz} by the following improved formulas:
\begin{equation}\label{equ:refined}
	\begin{split}
		n_v^A = n_{vx} + n_{vy} + n_{vz} + n_{vv},&\ E_v^A = n_{vv}  / n_v^A;\\
		n_x^A = n_{xx} + n_{xy} + n_{xz} + n_{xv},&\ E_x^A = (n_{xx} + n_{xy} + n_{xz}) / n_x^A;\\
		n_y^A = n_{yx} + n_{yy} + n_{yz} + n_{yv},&\ E_y^A = (n_{yx} + n_{yy} + n_{yz}) / n_y^A;\\
		n_z^A = n_{zx} + n_{zy} + n_{zz} + n_{zv},&\ E_z^A = (n_{zx} + n_{zy} + n_{zz}) / n_z^A.
	\end{split}
\end{equation}
Note that these values of $n_l^A$ and $E_l^A$ can be directly observed in the experiment.
Consequently, we have the following improved key length formula:
\begin{equation}\label{equ:refined_keylength}
	\begin{split}
		n= (n_0 + n_1) - n_1 H(e_1^{ph}) - f [n_v^A H(E_v^A)+n_x^A H(E_x^A)+n_y^A H(E_y^A)+n_z^A H(E_z^A)]  
		- 2\left(\log_2\frac{2}{\varepsilon_\text{cor}} - 2 \log_2\frac{1}{\sqrt{2}\varepsilon_\text{PA}\hat\varepsilon}\right),
	\end{split}
\end{equation}
where $n_0= n_{01}^v + n_{10}^v$ and $n_{01}^v$ ($n_{10}^v$) is the number of heralded time windows where Alice (Bob) decides a vacuum while Bob (Alice) decides a non vacuum intensity and he (she) has actually sent out vacuum~\cite{lo2005getting,chau2020security}.
In the case after AOPP, each of Alice's bit pair must contain one bit when she decides to send out a non-vacuum pulse and the other bit when she decides to send out vacuum.
Thus, all bit pairs passing the parity rejection of AOPP can be divided into three groups according to Alice's choice of her non-vacuum pulses in these pairs.
The number of bit pairs and the QBER of each group are $n_l^{\prime A}$ and $E_l^{\prime A}$ with $l\in\{x,y,z\}$.
Values of $n_l^{\prime A}$ and $E_l^{\prime A}$ are directly observed values of Alice. 
Expectedly, they are:
\begin{equation}\label{equ:refined_AOPP}
	\begin{split}
		n_x^{\prime A} = \sum_{r\in\{x,y,z\}}(n_{xr+vv}+n_{xv+vr}),&\ E_x^{\prime A} = \sum_{r\in\{x,y,z\}}n_{xr+vv} / n_x^{\prime A};\\
		n_y^{\prime A} = \sum_{r\in\{x,y,z\}}(n_{yr+vv}+n_{yv+vr}),&\ E_y^{\prime A} = \sum_{r\in\{x,y,z\}}n_{yr+vv} / n_y^{\prime A};\\
		n_z^{\prime A} = \sum_{r\in\{x,y,z\}}(n_{zr+vv}+n_{zv+vr}),&\ E_z^{\prime A} = \sum_{r\in\{x,y,z\}}n_{zr+vv} / n_z^{\prime A}.
	\end{split}
\end{equation}
Here, we use the notation ${lr+l^\prime r^\prime}$ for the bit pair in which one code bit comes from heralded $lr$-event and the other comes from heralded $l^\prime r^\prime$-event.
In an experiment, Alice does not need these formulas. 
She directly uses her observed values of $n_l^{\prime A}$ and $E_l^{\prime A}$ in the calculation of final key length. 
The formulas above are only useful in the numerical simulation for key length and optimization.
Consequently, we also have the following improved key length formula with AOPP:
\begin{equation}\label{equ:refined_keylength_AOPP}
	\begin{split}
		n^\prime= n_u^\prime [1 -  H(e_1^{\prime\prime ph})] - f [n_x^{\prime A} H(E_x^{\prime A})+n_y^{\prime A} H(E_y^{\prime A})+n_z^{\prime A} H(E_z^{\prime A})] 
	- 2\left(\log_2\frac{2}{\varepsilon_\text{cor}} - 2 \log_2\frac{1}{\sqrt{2}\varepsilon_\text{PA}\hat\varepsilon}\right),
	\end{split}
\end{equation}
where $e_1^{\prime\prime ph}=n_1^\prime e_1^{\prime ph}/n_u^\prime$ and $n_u^\prime$ is the number of untagged bits after post-selection of AOPP, dependent on the numbers of untagged bits before AOPP, $n_{01}$, $n_{10}$,  $n_{01}^v$ , $n_{10}^v$.

\subsection{Further improvement}

The secure key length can be further improved if we modify the protocols with refined analysis. 
For example, we modify the Protocol 1’ by the following Protocol M1’ with code-bit option $[y,z]$:\\
1. After quantum communication, they each announces the intensities used in those null time windows and those $|k_1-k_0|$ heralded time windows whose code bits are not paired with other bits in Alice’s odd-parity pairing. 
They also announce those heralded time windows when she/he has used intensity $\mu_x$.
Note: Given $k_1$ heralded time windows for code bits with bit value 1 and $k_0$ heralded time windows for code bits with bit value 0, Alice can only make $k=\min\{k_1, k_0\}$ odd-parity pairs in her AOPP, there are $|k_1-k_0|$ heralded time windows whose code bits are not paired with other bits. \\
2. They perform AOPP where Alice makes the odd-parity pairs, and do bit-flip error correction where Bob computes the positions of wrong bits, which means they will use Alice’s final bits for their shared final key. Bob keeps the positions of wrong bits as private information which will never be announced.\\
3. Alice announces the intensity of non-vacuum pulse  ($\mu_y$ or $\mu_z$) she has used in each of her AOPP pairs. (She does not announce in which time window she has used the non-vacuum intensity for her AOPP pair.)\\
4. They publicly announce the phase information of all heralded $xx$ windows. \\
5. Knowing the numbers of all kinds of events now, Bob takes decoy-state analysis with them. 
He obtains the lower bounds of $n_{10}^\prime$ and $n_{01}^\prime$ and the upper bound of $e_1^{\prime ph}$.  \\
6. Bob announces the final key length calculated by the following formula:
\begin{equation}\label{equ:keylength_tilde_n_Delta'}
	\tilde n^\prime = n_1^\prime [1 - H(e_1^{\prime ph})] - f n_t^\prime H(E_t^\prime) 
	- 2\left(\log_2\frac{2}{\varepsilon_\text{cor}} - 2 \log_2\frac{1}{\sqrt{2}\varepsilon_\text{PA}\hat\varepsilon}\right) - \Delta^\prime,
\end{equation}
where $\Delta^\prime = \log_2 \bar {n^\prime}$ and $\bar {n^\prime}$ is an upper bound of $n^\prime$. 
We can simply choose $\bar {n^\prime} = n_t^\prime$. 
There are obviously tighter bounds such as $n_t^\prime[1-f H(E_t^\prime)]$, but this changes the key length only by a negligibly small amount. 
We can also replace the term $f n_t^\prime H(E_t^\prime)$ in Eq.~\eqref{equ:keylength_tilde_n_Delta'} by $f [n_x^{\prime A} H(E_x^{\prime A})+n_y^{\prime A} H(E_y^{\prime A})+n_z^{\prime A} H(E_z^{\prime A})] $ to obtain a better result of key length.


Remark: Although the confidential numbers such as $n_{01}^\prime$, $n_{10}^\prime$ and $e_1^{\prime ph}$ are never announced by Bob, he has used them in his decoy-state analysis and the announced final key length is dependent on these confidential numbers. 
Therefore we deduct $\Delta^\prime$ bits in our key length formula Eq.~\eqref{equ:keylength_tilde_n_Delta'} according to Result 2 in Sec.~\ref{sec:decoy}.
This deduction $\Delta^\prime$ is a little smaller than $\Delta$ in Eq.~\eqref{equ:tilde_n_AOPP}.

We can also add a vacuum-related term to the key length formula in Eq.~\eqref{equ:keylength_tilde_n_Delta'} to further improve the key length:
\begin{equation}\label{equ:keylength_n0'}
	\tilde n^{\prime\prime} = n_0^\prime +\tilde n^\prime.
\end{equation}
We define a pair made by Alice in AOPP as a VA pair if both of Alice’s bits in this pair are from time windows when she has actually sent out vacuum. 
A bit post-selected from a VA pair is called a VA bit. 
Obviously, such a VA bit is an entirely private bit from Alice, and no one except Alice can have any information about it: both of Alice's bits in this VA pair correspond to identical vacuum pulses from her side, and thus people outside her lab can only know that these two bits are in odd parity but have no information on which one takes bit value 1 and which one takes bit value 0. 
Surely, $n_0^\prime$ is the number of VA pairs that pass thorough the parity check, i.e., the number of VA pairs with odd parity at Bob’s side.
It can be calculated by the following formula where all parameters are Bob’s observed values: 
\begin{equation}\label{equ:n0'}
	n_0^\prime= \zeta k_{\text{VA}} ,
\end{equation}
where $k_{\text{VA}}$ is the number of VA pairs in Alice’s AOPP before parity check and $\zeta$ is the surviving rate in the parity check to all VA pairs made in AOPP, which is just the VA pairs’ odd-parity rate at Bob’s side.
Asymptotically, there are 
\begin{equation}\label{equ:kVA}
	k_{\text{VA}} = r n_v^A q /p_v
\end{equation}
VA pairs among all those $k$ pairs made by Alice in AOPP, where  
$r=\min\{1, k_0/k_1\}$
and $q=\sum_{\alpha=y,z}p_\alpha e^{-\mu_\alpha}$.
Bob can verify $\zeta$ asymptotically by randomly pairing code bits from heralded time windows when Alice uses intensity $\mu_v$ and then checking the parity of his own bit values in each of those pairs.
Say, if he has made $m_0$ random pairs from those bits with Alice’s bit values 0, and then finds that among these $m_0$ pairs there are $m_d$ pairs taking odd parity of his own bit values, asymptotically he will obtain $\zeta = m_d/m_0$.

Remark B1: Here, ``his own bit values'' refers to Bob's original bit values before error-correction.

Remark B2: Since Bob computes the position of wrong bits in error correction, he knows Alice's bit values and thus knows which intensity Alice uses in each time window.

Remark B3: 
Alice’s announcement in Step 3 of protocol M1’ does not cause any information leakage to Alice’s secure bits because they are independent of which source ($\mu_y$ or $\mu_z$) Alice’s state comes from. 
Say, after parity check to those pairs made by Alice in AOPP, there are two kinds of untagged pairs: the VA pairs and those $n_1^\prime$ pairs when a single photon is actually sent out in each of corresponding heralded time windows. 
Bit values contributed by VA pairs are still entirely private even Alice makes the announcement in Step 3 above. 
Bit value contributed by any single-photon pair is entirely independent of which source ($\mu_y$ or $\mu_z$) Alice’s single photon has come from.


In the protocol above, with the amount $\Delta^\prime$ being deducted from the final key length, we can freely use all confidential observed numbers such as \textcolor{black}{$k_0$, $k_1$, $\zeta$, and $n_{lr}$ in calculation of the final key length.}
The finite-data-size effects can be easily taken to $n_0^\prime$ by adding statistical fluctuations to $k_{\text{VA}}$ and $\zeta$.

The improved key-length formula in Eq.~\eqref{equ:keylength_n0'} can also apply to the code-bit option $[z]$ and $[x,y,z]$. 
If option $[z]$ is chosen, we change $q$ into $p_z e^{-\mu_z}$ in Eq.~\eqref{equ:kVA} accordingly, and let them announce all those time windows when she/he has used intensity $\mu_x$ or $\mu_y$.
If option $[x,y,z]$ is chosen, the formula for $n_0^\prime$ is changed into
\begin{equation}\label{equ:n0'_xyz}
	n_0^\prime = \zeta k_{\text{VA}} - \delta_x,
\end{equation}
where $k_{\text{VA}}$ can be calculated by Eq.~\eqref{equ:kVA} with $q = \sum_{\alpha=x,y,z} p_\alpha e^{-\mu_\alpha}$, $\delta_x = \zeta_x k_x$.
Here, $k_x = r n_v^A p_x e^{-\mu_x} / p_v $ is the number of VA pairs from time windows when Alice has chosen intensity $\mu_x$ in one time window, and $\zeta_x$ is the proportion of pairs that the parity of Bob’s own bit values is odd and he has used intensity $\mu_x$ in all these $k_x$ VA pairs.
Similar to verification of the value of $\zeta$ above, Bob can verify $\zeta_x$ by making random pairs from bits with Alice's bit values 0 and observing the proportion of pairs that the parity of Bob's bits is odd and Bob has used intensity $\mu_x$ in these random pairs.
Say, if he has made $m_0^\prime$ random pairs from those bits with Alice’s bit values 0, and then finds that among these $m_0^\prime$ pairs there are $m_d^\prime$ pairs in which the parity of his own bit values is odd and he has used intensity $\mu_x$, asymptotically he will obtain $\zeta_x = m_d^\prime/m_0^\prime$.

If $p_z$ is set to be 0, the code-bit option $[x, y, z]$ and $[y,z]$ are equivalent to a 3-intensity protocol with option $[x,y]$ and $[y]$, respectively. 
Using the 3-intensity protocol with option $[y]$ or the 4-intensity protocol with option $[z]$, Step 3 in Protocol M1’ is not necessary.

\subsection{Scanning of $\mean{S_{vv}}$}
The bounds of $n_1$ and $e_1^{ph}$ depend on the value of $\mean{S_{vv}}$, and thus the key length $\mathcal N$ is a function of $\mean{S_{vv}}$, i.e. $\mathcal N(\mean{S_{vv}})$.
Surely, we can use the following more efficient key-length formula 
\begin{equation}\label{fffc}
	n_{\text{scan}} = \min\limits_{\mean{S_{vv}}} \mathcal N(\mean{S_{vv}}),
\end{equation}
i.e., by scanning  $\mean{S_{vv}}$ in its possible range for the worst-case result of $\mathcal N$ instead of taking worst-case separately for $n_1$ and $e_1^{ph}$, to improve the non-asymptotic key rate a little bit. 
Also, we can use similar scanning method in the key length formula after AOPP.
Here, $\mathcal N$ can be any key length function $\tilde n$, $n^\prime$, $\tilde n^\prime$, or $\tilde n^{\prime\prime}$ in Eqs.~\eqref{equ:tilde_n}~\eqref{equ:tilde_n_AOPP}~\eqref{equ:refined_keylength_AOPP}~\eqref{equ:keylength_tilde_n_Delta'}~\eqref{equ:keylength_n0'}.

\bibliography{refs}

\begin{thebibliography}{61}%
\makeatletter
\providecommand \@ifxundefined [1]{%
 \@ifx{#1\undefined}
}%
\providecommand \@ifnum [1]{%
 \ifnum #1\expandafter \@firstoftwo
 \else \expandafter \@secondoftwo
 \fi
}%
\providecommand \@ifx [1]{%
 \ifx #1\expandafter \@firstoftwo
 \else \expandafter \@secondoftwo
 \fi
}%
\providecommand \natexlab [1]{#1}%
\providecommand \enquote  [1]{``#1''}%
\providecommand \bibnamefont  [1]{#1}%
\providecommand \bibfnamefont [1]{#1}%
\providecommand \citenamefont [1]{#1}%
\providecommand \href@noop [0]{\@secondoftwo}%
\providecommand \href [0]{\begingroup \@sanitize@url \@href}%
\providecommand \@href[1]{\@@startlink{#1}\@@href}%
\providecommand \@@href[1]{\endgroup#1\@@endlink}%
\providecommand \@sanitize@url [0]{\catcode `\\12\catcode `\$12\catcode
  `\&12\catcode `\#12\catcode `\^12\catcode `\_12\catcode `\%12\relax}%
\providecommand \@@startlink[1]{}%
\providecommand \@@endlink[0]{}%
\providecommand \url  [0]{\begingroup\@sanitize@url \@url }%
\providecommand \@url [1]{\endgroup\@href {#1}{\urlprefix }}%
\providecommand \urlprefix  [0]{URL }%
\providecommand \Eprint [0]{\href }%
\providecommand \doibase [0]{http://dx.doi.org/}%
\providecommand \selectlanguage [0]{\@gobble}%
\providecommand \bibinfo  [0]{\@secondoftwo}%
\providecommand \bibfield  [0]{\@secondoftwo}%
\providecommand \translation [1]{[#1]}%
\providecommand \BibitemOpen [0]{}%
\providecommand \bibitemStop [0]{}%
\providecommand \bibitemNoStop [0]{.\EOS\space}%
\providecommand \EOS [0]{\spacefactor3000\relax}%
\providecommand \BibitemShut  [1]{\csname bibitem#1\endcsname}%
\let\auto@bib@innerbib\@empty
\bibitem [{\citenamefont {Bennett}\ and\ \citenamefont
  {Gilles}(1984)}]{bennett1984quantum}%
  \BibitemOpen
  \bibfield  {author} {\bibinfo {author} {\bibfnamefont {C.~H.}\ \bibnamefont
  {Bennett}}\ and\ \bibinfo {author} {\bibfnamefont {B.}~\bibnamefont
  {Gilles}},\ }in\ \href@noop {} {\emph {\bibinfo {booktitle} {Proceedings of
  the IEEE International\ Conference on Computers, Systems, and Signal
  Processing}}}\ (\bibinfo {year} {1984})\ pp.\ \bibinfo {pages}
  {175--179}\BibitemShut {NoStop}%
\bibitem [{\citenamefont {Lo}\ and\ \citenamefont
  {Chau}(1999)}]{lo1999unconditional}%
  \BibitemOpen
  \bibfield  {author} {\bibinfo {author} {\bibfnamefont {H.-K.}\ \bibnamefont
  {Lo}}\ and\ \bibinfo {author} {\bibfnamefont {H.~F.}\ \bibnamefont {Chau}},\
  }\href@noop {} {\bibfield  {journal} {\bibinfo  {journal} {science}\ }\textbf
  {\bibinfo {volume} {283}},\ \bibinfo {pages} {2050} (\bibinfo {year}
  {1999})}\BibitemShut {NoStop}%
\bibitem [{\citenamefont {Shor}\ and\ \citenamefont
  {Preskill}(2000)}]{shor2000simple}%
  \BibitemOpen
  \bibfield  {author} {\bibinfo {author} {\bibfnamefont {P.~W.}\ \bibnamefont
  {Shor}}\ and\ \bibinfo {author} {\bibfnamefont {J.}~\bibnamefont
  {Preskill}},\ }\href@noop {} {\bibfield  {journal} {\bibinfo  {journal}
  {Physical Review Letters}\ }\textbf {\bibinfo {volume} {85}},\ \bibinfo
  {pages} {441} (\bibinfo {year} {2000})}\BibitemShut {NoStop}%
\bibitem [{\citenamefont {Renner}(2008)}]{renner2008security}%
  \BibitemOpen
  \bibfield  {author} {\bibinfo {author} {\bibfnamefont {R.}~\bibnamefont
  {Renner}},\ }\href@noop {} {\bibfield  {journal} {\bibinfo  {journal}
  {International Journal of Quantum Information}\ }\textbf {\bibinfo {volume}
  {6}},\ \bibinfo {pages} {1} (\bibinfo {year} {2008})}\BibitemShut {NoStop}%
\bibitem [{\citenamefont {Scarani}\ \emph {et~al.}(2009)\citenamefont
  {Scarani}, \citenamefont {Bechmann-Pasquinucci}, \citenamefont {Cerf},
  \citenamefont {Du{\v{s}}ek}, \citenamefont {L{\"u}tkenhaus},\ and\
  \citenamefont {Peev}}]{scarani2009security}%
  \BibitemOpen
  \bibfield  {author} {\bibinfo {author} {\bibfnamefont {V.}~\bibnamefont
  {Scarani}}, \bibinfo {author} {\bibfnamefont {H.}~\bibnamefont
  {Bechmann-Pasquinucci}}, \bibinfo {author} {\bibfnamefont {N.~J.}\
  \bibnamefont {Cerf}}, \bibinfo {author} {\bibfnamefont {M.}~\bibnamefont
  {Du{\v{s}}ek}}, \bibinfo {author} {\bibfnamefont {N.}~\bibnamefont
  {L{\"u}tkenhaus}}, \ and\ \bibinfo {author} {\bibfnamefont {M.}~\bibnamefont
  {Peev}},\ }\href@noop {} {\bibfield  {journal} {\bibinfo  {journal} {Reviews
  of modern physics}\ }\textbf {\bibinfo {volume} {81}},\ \bibinfo {pages}
  {1301} (\bibinfo {year} {2009})}\BibitemShut {NoStop}%
\bibitem [{\citenamefont {Tomamichel}\ \emph {et~al.}(2012)\citenamefont
  {Tomamichel}, \citenamefont {Lim}, \citenamefont {Gisin},\ and\ \citenamefont
  {Renner}}]{tomamichel2012tight}%
  \BibitemOpen
  \bibfield  {author} {\bibinfo {author} {\bibfnamefont {M.}~\bibnamefont
  {Tomamichel}}, \bibinfo {author} {\bibfnamefont {C.~C.~W.}\ \bibnamefont
  {Lim}}, \bibinfo {author} {\bibfnamefont {N.}~\bibnamefont {Gisin}}, \ and\
  \bibinfo {author} {\bibfnamefont {R.}~\bibnamefont {Renner}},\ }\href@noop {}
  {\bibfield  {journal} {\bibinfo  {journal} {Nature Communications}\ }\textbf
  {\bibinfo {volume} {3}},\ \bibinfo {pages} {634} (\bibinfo {year}
  {2012})}\BibitemShut {NoStop}%
\bibitem [{\citenamefont {Xu}\ \emph {et~al.}(2020{\natexlab{a}})\citenamefont
  {Xu}, \citenamefont {Ma}, \citenamefont {Zhang}, \citenamefont {Lo},\ and\
  \citenamefont {Pan}}]{xu2020secure}%
  \BibitemOpen
  \bibfield  {author} {\bibinfo {author} {\bibfnamefont {F.}~\bibnamefont
  {Xu}}, \bibinfo {author} {\bibfnamefont {X.}~\bibnamefont {Ma}}, \bibinfo
  {author} {\bibfnamefont {Q.}~\bibnamefont {Zhang}}, \bibinfo {author}
  {\bibfnamefont {H.-K.}\ \bibnamefont {Lo}}, \ and\ \bibinfo {author}
  {\bibfnamefont {J.-W.}\ \bibnamefont {Pan}},\ }\href@noop {} {\bibfield
  {journal} {\bibinfo  {journal} {Reviews of Modern Physics}\ }\textbf
  {\bibinfo {volume} {92}},\ \bibinfo {pages} {025002} (\bibinfo {year}
  {2020}{\natexlab{a}})}\BibitemShut {NoStop}%
\bibitem [{\citenamefont {Pirandola}\ \emph {et~al.}(2020)\citenamefont
  {Pirandola}, \citenamefont {Andersen}, \citenamefont {Banchi}, \citenamefont
  {Berta}, \citenamefont {Bunandar}, \citenamefont {Colbeck}, \citenamefont
  {Englund}, \citenamefont {Gehring}, \citenamefont {Lupo}, \citenamefont
  {Ottaviani} \emph {et~al.}}]{pirandola2020advances}%
  \BibitemOpen
  \bibfield  {author} {\bibinfo {author} {\bibfnamefont {S.}~\bibnamefont
  {Pirandola}}, \bibinfo {author} {\bibfnamefont {U.~L.}\ \bibnamefont
  {Andersen}}, \bibinfo {author} {\bibfnamefont {L.}~\bibnamefont {Banchi}},
  \bibinfo {author} {\bibfnamefont {M.}~\bibnamefont {Berta}}, \bibinfo
  {author} {\bibfnamefont {D.}~\bibnamefont {Bunandar}}, \bibinfo {author}
  {\bibfnamefont {R.}~\bibnamefont {Colbeck}}, \bibinfo {author} {\bibfnamefont
  {D.}~\bibnamefont {Englund}}, \bibinfo {author} {\bibfnamefont
  {T.}~\bibnamefont {Gehring}}, \bibinfo {author} {\bibfnamefont
  {C.}~\bibnamefont {Lupo}}, \bibinfo {author} {\bibfnamefont {C.}~\bibnamefont
  {Ottaviani}},  \emph {et~al.},\ }\href@noop {} {\bibfield  {journal}
  {\bibinfo  {journal} {Advances in Optics and Photonics}\ }\textbf {\bibinfo
  {volume} {12}},\ \bibinfo {pages} {1012} (\bibinfo {year}
  {2020})}\BibitemShut {NoStop}%
\bibitem [{\citenamefont {Hwang}(2003)}]{hwang2003quantum}%
  \BibitemOpen
  \bibfield  {author} {\bibinfo {author} {\bibfnamefont {W.-Y.}\ \bibnamefont
  {Hwang}},\ }\href@noop {} {\bibfield  {journal} {\bibinfo  {journal}
  {Physical Review Letters}\ }\textbf {\bibinfo {volume} {91}},\ \bibinfo
  {pages} {057901} (\bibinfo {year} {2003})}\BibitemShut {NoStop}%
\bibitem [{\citenamefont {Wang}(2005)}]{wang2005beating}%
  \BibitemOpen
  \bibfield  {author} {\bibinfo {author} {\bibfnamefont {X.-B.}\ \bibnamefont
  {Wang}},\ }\href@noop {} {\bibfield  {journal} {\bibinfo  {journal} {Physical
  Review Letters}\ }\textbf {\bibinfo {volume} {94}},\ \bibinfo {pages}
  {230503} (\bibinfo {year} {2005})}\BibitemShut {NoStop}%
\bibitem [{\citenamefont {Lo}\ \emph {et~al.}(2005)\citenamefont {Lo},
  \citenamefont {Ma},\ and\ \citenamefont {Chen}}]{lo2005decoy}%
  \BibitemOpen
  \bibfield  {author} {\bibinfo {author} {\bibfnamefont {H.-K.}\ \bibnamefont
  {Lo}}, \bibinfo {author} {\bibfnamefont {X.}~\bibnamefont {Ma}}, \ and\
  \bibinfo {author} {\bibfnamefont {K.}~\bibnamefont {Chen}},\ }\href@noop {}
  {\bibfield  {journal} {\bibinfo  {journal} {Physical Review Letters}\
  }\textbf {\bibinfo {volume} {94}},\ \bibinfo {pages} {230504} (\bibinfo
  {year} {2005})}\BibitemShut {NoStop}%
\bibitem [{\citenamefont {Lim}\ \emph {et~al.}(2014)\citenamefont {Lim},
  \citenamefont {Curty}, \citenamefont {Walenta}, \citenamefont {Xu},\ and\
  \citenamefont {Zbinden}}]{lim2014concise}%
  \BibitemOpen
  \bibfield  {author} {\bibinfo {author} {\bibfnamefont {C.~C.~W.}\
  \bibnamefont {Lim}}, \bibinfo {author} {\bibfnamefont {M.}~\bibnamefont
  {Curty}}, \bibinfo {author} {\bibfnamefont {N.}~\bibnamefont {Walenta}},
  \bibinfo {author} {\bibfnamefont {F.}~\bibnamefont {Xu}}, \ and\ \bibinfo
  {author} {\bibfnamefont {H.}~\bibnamefont {Zbinden}},\ }\href@noop {}
  {\bibfield  {journal} {\bibinfo  {journal} {Physical Review A}\ }\textbf
  {\bibinfo {volume} {89}},\ \bibinfo {pages} {022307} (\bibinfo {year}
  {2014})}\BibitemShut {NoStop}%
\bibitem [{\citenamefont {Lo}\ \emph {et~al.}(2012)\citenamefont {Lo},
  \citenamefont {Curty},\ and\ \citenamefont {Qi}}]{lo2012measurement}%
  \BibitemOpen
  \bibfield  {author} {\bibinfo {author} {\bibfnamefont {H.-K.}\ \bibnamefont
  {Lo}}, \bibinfo {author} {\bibfnamefont {M.}~\bibnamefont {Curty}}, \ and\
  \bibinfo {author} {\bibfnamefont {B.}~\bibnamefont {Qi}},\ }\href@noop {}
  {\bibfield  {journal} {\bibinfo  {journal} {Physical Review Letters}\
  }\textbf {\bibinfo {volume} {108}},\ \bibinfo {pages} {130503} (\bibinfo
  {year} {2012})}\BibitemShut {NoStop}%
\bibitem [{\citenamefont {Braunstein}\ and\ \citenamefont
  {Pirandola}(2012)}]{braunstein2012side}%
  \BibitemOpen
  \bibfield  {author} {\bibinfo {author} {\bibfnamefont {S.~L.}\ \bibnamefont
  {Braunstein}}\ and\ \bibinfo {author} {\bibfnamefont {S.}~\bibnamefont
  {Pirandola}},\ }\href@noop {} {\bibfield  {journal} {\bibinfo  {journal}
  {Physical Review Letters}\ }\textbf {\bibinfo {volume} {108}},\ \bibinfo
  {pages} {130502} (\bibinfo {year} {2012})}\BibitemShut {NoStop}%
\bibitem [{\citenamefont {Tamaki}\ \emph {et~al.}(2012)\citenamefont {Tamaki},
  \citenamefont {Lo}, \citenamefont {Fung},\ and\ \citenamefont
  {Qi}}]{tamaki2012phase}%
  \BibitemOpen
  \bibfield  {author} {\bibinfo {author} {\bibfnamefont {K.}~\bibnamefont
  {Tamaki}}, \bibinfo {author} {\bibfnamefont {H.-K.}\ \bibnamefont {Lo}},
  \bibinfo {author} {\bibfnamefont {C.-H.~F.}\ \bibnamefont {Fung}}, \ and\
  \bibinfo {author} {\bibfnamefont {B.}~\bibnamefont {Qi}},\ }\href@noop {}
  {\bibfield  {journal} {\bibinfo  {journal} {Physical Review A}\ }\textbf
  {\bibinfo {volume} {85}},\ \bibinfo {pages} {042307} (\bibinfo {year}
  {2012})}\BibitemShut {NoStop}%
\bibitem [{\citenamefont {Wang}(2013)}]{wang2013three}%
  \BibitemOpen
  \bibfield  {author} {\bibinfo {author} {\bibfnamefont {X.-B.}\ \bibnamefont
  {Wang}},\ }\href@noop {} {\bibfield  {journal} {\bibinfo  {journal} {Physical
  Review A}\ }\textbf {\bibinfo {volume} {87}},\ \bibinfo {pages} {012320}
  (\bibinfo {year} {2013})}\BibitemShut {NoStop}%
\bibitem [{\citenamefont {Curty}\ \emph {et~al.}(2014)\citenamefont {Curty},
  \citenamefont {Xu}, \citenamefont {Cui}, \citenamefont {Lim}, \citenamefont
  {Tamaki},\ and\ \citenamefont {Lo}}]{curty2014finite}%
  \BibitemOpen
  \bibfield  {author} {\bibinfo {author} {\bibfnamefont {M.}~\bibnamefont
  {Curty}}, \bibinfo {author} {\bibfnamefont {F.}~\bibnamefont {Xu}}, \bibinfo
  {author} {\bibfnamefont {W.}~\bibnamefont {Cui}}, \bibinfo {author}
  {\bibfnamefont {C.~C.~W.}\ \bibnamefont {Lim}}, \bibinfo {author}
  {\bibfnamefont {K.}~\bibnamefont {Tamaki}}, \ and\ \bibinfo {author}
  {\bibfnamefont {H.-K.}\ \bibnamefont {Lo}},\ }\href@noop {} {\bibfield
  {journal} {\bibinfo  {journal} {Nature communications}\ }\textbf {\bibinfo
  {volume} {5}} (\bibinfo {year} {2014})}\BibitemShut {NoStop}%
\bibitem [{\citenamefont {Xu}\ \emph {et~al.}(2014)\citenamefont {Xu},
  \citenamefont {Xu},\ and\ \citenamefont {Lo}}]{xu2014protocol}%
  \BibitemOpen
  \bibfield  {author} {\bibinfo {author} {\bibfnamefont {F.}~\bibnamefont
  {Xu}}, \bibinfo {author} {\bibfnamefont {H.}~\bibnamefont {Xu}}, \ and\
  \bibinfo {author} {\bibfnamefont {H.-K.}\ \bibnamefont {Lo}},\ }\href@noop {}
  {\bibfield  {journal} {\bibinfo  {journal} {Physical Review A}\ }\textbf
  {\bibinfo {volume} {89}},\ \bibinfo {pages} {052333} (\bibinfo {year}
  {2014})}\BibitemShut {NoStop}%
\bibitem [{\citenamefont {Yu}\ \emph {et~al.}(2015)\citenamefont {Yu},
  \citenamefont {Zhou},\ and\ \citenamefont {Wang}}]{yu2015statistical}%
  \BibitemOpen
  \bibfield  {author} {\bibinfo {author} {\bibfnamefont {Z.-W.}\ \bibnamefont
  {Yu}}, \bibinfo {author} {\bibfnamefont {Y.-H.}\ \bibnamefont {Zhou}}, \ and\
  \bibinfo {author} {\bibfnamefont {X.-B.}\ \bibnamefont {Wang}},\ }\href@noop
  {} {\bibfield  {journal} {\bibinfo  {journal} {Physical Review A}\ }\textbf
  {\bibinfo {volume} {91}},\ \bibinfo {pages} {032318} (\bibinfo {year}
  {2015})}\BibitemShut {NoStop}%
\bibitem [{\citenamefont {Zhou}\ \emph {et~al.}(2016)\citenamefont {Zhou},
  \citenamefont {Yu},\ and\ \citenamefont {Wang}}]{zhou2016making}%
  \BibitemOpen
  \bibfield  {author} {\bibinfo {author} {\bibfnamefont {Y.-H.}\ \bibnamefont
  {Zhou}}, \bibinfo {author} {\bibfnamefont {Z.-W.}\ \bibnamefont {Yu}}, \ and\
  \bibinfo {author} {\bibfnamefont {X.-B.}\ \bibnamefont {Wang}},\ }\href@noop
  {} {\bibfield  {journal} {\bibinfo  {journal} {Physical Review A}\ }\textbf
  {\bibinfo {volume} {93}},\ \bibinfo {pages} {042324} (\bibinfo {year}
  {2016})}\BibitemShut {NoStop}%
\bibitem [{\citenamefont {Hu}\ \emph {et~al.}(2021)\citenamefont {Hu},
  \citenamefont {Jiang}, \citenamefont {Yu},\ and\ \citenamefont
  {Wang}}]{hu2021practical}%
  \BibitemOpen
  \bibfield  {author} {\bibinfo {author} {\bibfnamefont {X.-L.}\ \bibnamefont
  {Hu}}, \bibinfo {author} {\bibfnamefont {C.}~\bibnamefont {Jiang}}, \bibinfo
  {author} {\bibfnamefont {Z.-W.}\ \bibnamefont {Yu}}, \ and\ \bibinfo {author}
  {\bibfnamefont {X.-B.}\ \bibnamefont {Wang}},\ }\href@noop {} {\bibfield
  {journal} {\bibinfo  {journal} {Advanced Quantum Technologies}\ }\textbf
  {\bibinfo {volume} {4}},\ \bibinfo {pages} {2100069} (\bibinfo {year}
  {2021})}\BibitemShut {NoStop}%
\bibitem [{\citenamefont {Jiang}\ \emph
  {et~al.}(2021{\natexlab{a}})\citenamefont {Jiang}, \citenamefont {Yu},
  \citenamefont {Hu},\ and\ \citenamefont {Wang}}]{jiang2021higher}%
  \BibitemOpen
  \bibfield  {author} {\bibinfo {author} {\bibfnamefont {C.}~\bibnamefont
  {Jiang}}, \bibinfo {author} {\bibfnamefont {Z.-W.}\ \bibnamefont {Yu}},
  \bibinfo {author} {\bibfnamefont {X.-L.}\ \bibnamefont {Hu}}, \ and\ \bibinfo
  {author} {\bibfnamefont {X.-B.}\ \bibnamefont {Wang}},\ }\href@noop {}
  {\bibfield  {journal} {\bibinfo  {journal} {Physical Review A}\ }\textbf
  {\bibinfo {volume} {103}},\ \bibinfo {pages} {012402} (\bibinfo {year}
  {2021}{\natexlab{a}})}\BibitemShut {NoStop}%
\bibitem [{\citenamefont {Rubenok}\ \emph {et~al.}(2013)\citenamefont
  {Rubenok}, \citenamefont {Slater}, \citenamefont {Chan}, \citenamefont
  {Lucio-Martinez},\ and\ \citenamefont {Tittel}}]{rubenok2013real}%
  \BibitemOpen
  \bibfield  {author} {\bibinfo {author} {\bibfnamefont {A.}~\bibnamefont
  {Rubenok}}, \bibinfo {author} {\bibfnamefont {J.~A.}\ \bibnamefont {Slater}},
  \bibinfo {author} {\bibfnamefont {P.}~\bibnamefont {Chan}}, \bibinfo {author}
  {\bibfnamefont {I.}~\bibnamefont {Lucio-Martinez}}, \ and\ \bibinfo {author}
  {\bibfnamefont {W.}~\bibnamefont {Tittel}},\ }\href@noop {} {\bibfield
  {journal} {\bibinfo  {journal} {Physical Review Letters}\ }\textbf {\bibinfo
  {volume} {111}},\ \bibinfo {pages} {130501} (\bibinfo {year}
  {2013})}\BibitemShut {NoStop}%
\bibitem [{\citenamefont {Liu}\ \emph {et~al.}(2013)\citenamefont {Liu},
  \citenamefont {Chen}, \citenamefont {Wang}, \citenamefont {Liang},
  \citenamefont {Shentu}, \citenamefont {Wang}, \citenamefont {Cui},
  \citenamefont {Yin}, \citenamefont {Liu}, \citenamefont {Li} \emph
  {et~al.}}]{liu2013experimental}%
  \BibitemOpen
  \bibfield  {author} {\bibinfo {author} {\bibfnamefont {Y.}~\bibnamefont
  {Liu}}, \bibinfo {author} {\bibfnamefont {T.-Y.}\ \bibnamefont {Chen}},
  \bibinfo {author} {\bibfnamefont {L.-J.}\ \bibnamefont {Wang}}, \bibinfo
  {author} {\bibfnamefont {H.}~\bibnamefont {Liang}}, \bibinfo {author}
  {\bibfnamefont {G.-L.}\ \bibnamefont {Shentu}}, \bibinfo {author}
  {\bibfnamefont {J.}~\bibnamefont {Wang}}, \bibinfo {author} {\bibfnamefont
  {K.}~\bibnamefont {Cui}}, \bibinfo {author} {\bibfnamefont {H.-L.}\
  \bibnamefont {Yin}}, \bibinfo {author} {\bibfnamefont {N.-L.}\ \bibnamefont
  {Liu}}, \bibinfo {author} {\bibfnamefont {L.}~\bibnamefont {Li}},  \emph
  {et~al.},\ }\href@noop {} {\bibfield  {journal} {\bibinfo  {journal}
  {Physical Review Letters}\ }\textbf {\bibinfo {volume} {111}},\ \bibinfo
  {pages} {130502} (\bibinfo {year} {2013})}\BibitemShut {NoStop}%
\bibitem [{\citenamefont {Yin}\ \emph {et~al.}(2016)\citenamefont {Yin},
  \citenamefont {Chen}, \citenamefont {Yu}, \citenamefont {Liu}, \citenamefont
  {You}, \citenamefont {Zhou}, \citenamefont {Chen}, \citenamefont {Mao},
  \citenamefont {Huang}, \citenamefont {Zhang} \emph
  {et~al.}}]{yin2016measurement}%
  \BibitemOpen
  \bibfield  {author} {\bibinfo {author} {\bibfnamefont {H.-L.}\ \bibnamefont
  {Yin}}, \bibinfo {author} {\bibfnamefont {T.-Y.}\ \bibnamefont {Chen}},
  \bibinfo {author} {\bibfnamefont {Z.-W.}\ \bibnamefont {Yu}}, \bibinfo
  {author} {\bibfnamefont {H.}~\bibnamefont {Liu}}, \bibinfo {author}
  {\bibfnamefont {L.-X.}\ \bibnamefont {You}}, \bibinfo {author} {\bibfnamefont
  {Y.-H.}\ \bibnamefont {Zhou}}, \bibinfo {author} {\bibfnamefont {S.-J.}\
  \bibnamefont {Chen}}, \bibinfo {author} {\bibfnamefont {Y.}~\bibnamefont
  {Mao}}, \bibinfo {author} {\bibfnamefont {M.-Q.}\ \bibnamefont {Huang}},
  \bibinfo {author} {\bibfnamefont {W.-J.}\ \bibnamefont {Zhang}},  \emph
  {et~al.},\ }\href@noop {} {\bibfield  {journal} {\bibinfo  {journal}
  {Physical Review Letters}\ }\textbf {\bibinfo {volume} {117}},\ \bibinfo
  {pages} {190501} (\bibinfo {year} {2016})}\BibitemShut {NoStop}%
\bibitem [{\citenamefont {Comandar}\ \emph {et~al.}(2016)\citenamefont
  {Comandar}, \citenamefont {Lucamarini}, \citenamefont {Fr{\"o}hlich},
  \citenamefont {Dynes}, \citenamefont {Sharpe}, \citenamefont {Tam},
  \citenamefont {Yuan}, \citenamefont {Penty},\ and\ \citenamefont
  {Shields}}]{comandar2016quantum}%
  \BibitemOpen
  \bibfield  {author} {\bibinfo {author} {\bibfnamefont {L.}~\bibnamefont
  {Comandar}}, \bibinfo {author} {\bibfnamefont {M.}~\bibnamefont
  {Lucamarini}}, \bibinfo {author} {\bibfnamefont {B.}~\bibnamefont
  {Fr{\"o}hlich}}, \bibinfo {author} {\bibfnamefont {J.}~\bibnamefont {Dynes}},
  \bibinfo {author} {\bibfnamefont {A.}~\bibnamefont {Sharpe}}, \bibinfo
  {author} {\bibfnamefont {S.-B.}\ \bibnamefont {Tam}}, \bibinfo {author}
  {\bibfnamefont {Z.}~\bibnamefont {Yuan}}, \bibinfo {author} {\bibfnamefont
  {R.}~\bibnamefont {Penty}}, \ and\ \bibinfo {author} {\bibfnamefont
  {A.}~\bibnamefont {Shields}},\ }\href@noop {} {\bibfield  {journal} {\bibinfo
   {journal} {Nature Photonics}\ }\textbf {\bibinfo {volume} {10}},\ \bibinfo
  {pages} {312} (\bibinfo {year} {2016})}\BibitemShut {NoStop}%
\bibitem [{\citenamefont {Wang}\ \emph {et~al.}(2017)\citenamefont {Wang},
  \citenamefont {Yin}, \citenamefont {Wang}, \citenamefont {Chen},
  \citenamefont {Guo},\ and\ \citenamefont {Han}}]{wang2017measurement}%
  \BibitemOpen
  \bibfield  {author} {\bibinfo {author} {\bibfnamefont {C.}~\bibnamefont
  {Wang}}, \bibinfo {author} {\bibfnamefont {Z.-Q.}\ \bibnamefont {Yin}},
  \bibinfo {author} {\bibfnamefont {S.}~\bibnamefont {Wang}}, \bibinfo {author}
  {\bibfnamefont {W.}~\bibnamefont {Chen}}, \bibinfo {author} {\bibfnamefont
  {G.-C.}\ \bibnamefont {Guo}}, \ and\ \bibinfo {author} {\bibfnamefont
  {Z.-F.}\ \bibnamefont {Han}},\ }\href@noop {} {\bibfield  {journal} {\bibinfo
   {journal} {Optica}\ }\textbf {\bibinfo {volume} {4}},\ \bibinfo {pages}
  {1016} (\bibinfo {year} {2017})}\BibitemShut {NoStop}%
\bibitem [{\citenamefont {Semenenko}\ \emph {et~al.}(2020)\citenamefont
  {Semenenko}, \citenamefont {Sibson}, \citenamefont {Hart}, \citenamefont
  {Thompson}, \citenamefont {Rarity},\ and\ \citenamefont
  {Erven}}]{semenenko2020chip}%
  \BibitemOpen
  \bibfield  {author} {\bibinfo {author} {\bibfnamefont {H.}~\bibnamefont
  {Semenenko}}, \bibinfo {author} {\bibfnamefont {P.}~\bibnamefont {Sibson}},
  \bibinfo {author} {\bibfnamefont {A.}~\bibnamefont {Hart}}, \bibinfo {author}
  {\bibfnamefont {M.~G.}\ \bibnamefont {Thompson}}, \bibinfo {author}
  {\bibfnamefont {J.~G.}\ \bibnamefont {Rarity}}, \ and\ \bibinfo {author}
  {\bibfnamefont {C.}~\bibnamefont {Erven}},\ }\href@noop {} {\bibfield
  {journal} {\bibinfo  {journal} {Optica}\ }\textbf {\bibinfo {volume} {7}},\
  \bibinfo {pages} {238} (\bibinfo {year} {2020})}\BibitemShut {NoStop}%
\bibitem [{\citenamefont {Wei}\ \emph {et~al.}(2020)\citenamefont {Wei},
  \citenamefont {Li}, \citenamefont {Tan}, \citenamefont {Li}, \citenamefont
  {Min}, \citenamefont {Zhang}, \citenamefont {Li}, \citenamefont {You},
  \citenamefont {Wang}, \citenamefont {Jiang} \emph {et~al.}}]{wei2020high}%
  \BibitemOpen
  \bibfield  {author} {\bibinfo {author} {\bibfnamefont {K.}~\bibnamefont
  {Wei}}, \bibinfo {author} {\bibfnamefont {W.}~\bibnamefont {Li}}, \bibinfo
  {author} {\bibfnamefont {H.}~\bibnamefont {Tan}}, \bibinfo {author}
  {\bibfnamefont {Y.}~\bibnamefont {Li}}, \bibinfo {author} {\bibfnamefont
  {H.}~\bibnamefont {Min}}, \bibinfo {author} {\bibfnamefont {W.-J.}\
  \bibnamefont {Zhang}}, \bibinfo {author} {\bibfnamefont {H.}~\bibnamefont
  {Li}}, \bibinfo {author} {\bibfnamefont {L.}~\bibnamefont {You}}, \bibinfo
  {author} {\bibfnamefont {Z.}~\bibnamefont {Wang}}, \bibinfo {author}
  {\bibfnamefont {X.}~\bibnamefont {Jiang}},  \emph {et~al.},\ }\href@noop {}
  {\bibfield  {journal} {\bibinfo  {journal} {Physical Review X}\ }\textbf
  {\bibinfo {volume} {10}},\ \bibinfo {pages} {031030} (\bibinfo {year}
  {2020})}\BibitemShut {NoStop}%
\bibitem [{\citenamefont {Zheng}\ \emph {et~al.}(2021)\citenamefont {Zheng},
  \citenamefont {Zhang}, \citenamefont {Ge}, \citenamefont {Lu}, \citenamefont
  {He}, \citenamefont {Chen}, \citenamefont {Qu}, \citenamefont {Zhang},
  \citenamefont {Cai}, \citenamefont {Lu} \emph
  {et~al.}}]{zheng2021heterogeneously}%
  \BibitemOpen
  \bibfield  {author} {\bibinfo {author} {\bibfnamefont {X.}~\bibnamefont
  {Zheng}}, \bibinfo {author} {\bibfnamefont {P.}~\bibnamefont {Zhang}},
  \bibinfo {author} {\bibfnamefont {R.}~\bibnamefont {Ge}}, \bibinfo {author}
  {\bibfnamefont {L.}~\bibnamefont {Lu}}, \bibinfo {author} {\bibfnamefont
  {G.}~\bibnamefont {He}}, \bibinfo {author} {\bibfnamefont {Q.}~\bibnamefont
  {Chen}}, \bibinfo {author} {\bibfnamefont {F.}~\bibnamefont {Qu}}, \bibinfo
  {author} {\bibfnamefont {L.}~\bibnamefont {Zhang}}, \bibinfo {author}
  {\bibfnamefont {X.}~\bibnamefont {Cai}}, \bibinfo {author} {\bibfnamefont
  {Y.}~\bibnamefont {Lu}},  \emph {et~al.},\ }\href@noop {} {\bibfield
  {journal} {\bibinfo  {journal} {Advanced Photonics}\ }\textbf {\bibinfo
  {volume} {3}},\ \bibinfo {pages} {055002} (\bibinfo {year}
  {2021})}\BibitemShut {NoStop}%
\bibitem [{\citenamefont {Boaron}\ \emph {et~al.}(2018)\citenamefont {Boaron},
  \citenamefont {Boso}, \citenamefont {Rusca}, \citenamefont {Vulliez},
  \citenamefont {Autebert}, \citenamefont {Caloz}, \citenamefont {Perrenoud},
  \citenamefont {Gras}, \citenamefont {Bussi{\`e}res}, \citenamefont {Li} \emph
  {et~al.}}]{boaron2018secure}%
  \BibitemOpen
  \bibfield  {author} {\bibinfo {author} {\bibfnamefont {A.}~\bibnamefont
  {Boaron}}, \bibinfo {author} {\bibfnamefont {G.}~\bibnamefont {Boso}},
  \bibinfo {author} {\bibfnamefont {D.}~\bibnamefont {Rusca}}, \bibinfo
  {author} {\bibfnamefont {C.}~\bibnamefont {Vulliez}}, \bibinfo {author}
  {\bibfnamefont {C.}~\bibnamefont {Autebert}}, \bibinfo {author}
  {\bibfnamefont {M.}~\bibnamefont {Caloz}}, \bibinfo {author} {\bibfnamefont
  {M.}~\bibnamefont {Perrenoud}}, \bibinfo {author} {\bibfnamefont
  {G.}~\bibnamefont {Gras}}, \bibinfo {author} {\bibfnamefont {F.}~\bibnamefont
  {Bussi{\`e}res}}, \bibinfo {author} {\bibfnamefont {M.-J.}\ \bibnamefont
  {Li}},  \emph {et~al.},\ }\href@noop {} {\bibfield  {journal} {\bibinfo
  {journal} {Physical Review Letters}\ }\textbf {\bibinfo {volume} {121}},\
  \bibinfo {pages} {190502} (\bibinfo {year} {2018})}\BibitemShut {NoStop}%
\bibitem [{\citenamefont {Pirandola}\ \emph {et~al.}(2017)\citenamefont
  {Pirandola}, \citenamefont {Laurenza}, \citenamefont {Ottaviani},\ and\
  \citenamefont {Banchi}}]{pirandola2017fundamental}%
  \BibitemOpen
  \bibfield  {author} {\bibinfo {author} {\bibfnamefont {S.}~\bibnamefont
  {Pirandola}}, \bibinfo {author} {\bibfnamefont {R.}~\bibnamefont {Laurenza}},
  \bibinfo {author} {\bibfnamefont {C.}~\bibnamefont {Ottaviani}}, \ and\
  \bibinfo {author} {\bibfnamefont {L.}~\bibnamefont {Banchi}},\ }\href@noop {}
  {\bibfield  {journal} {\bibinfo  {journal} {Nature communications}\ }\textbf
  {\bibinfo {volume} {8}},\ \bibinfo {pages} {15043} (\bibinfo {year}
  {2017})}\BibitemShut {NoStop}%
\bibitem [{\citenamefont {Lucamarini}\ \emph {et~al.}(2018)\citenamefont
  {Lucamarini}, \citenamefont {Yuan}, \citenamefont {Dynes},\ and\
  \citenamefont {Shields}}]{lucamarini2018overcoming}%
  \BibitemOpen
  \bibfield  {author} {\bibinfo {author} {\bibfnamefont {M.}~\bibnamefont
  {Lucamarini}}, \bibinfo {author} {\bibfnamefont {Z.}~\bibnamefont {Yuan}},
  \bibinfo {author} {\bibfnamefont {J.}~\bibnamefont {Dynes}}, \ and\ \bibinfo
  {author} {\bibfnamefont {A.}~\bibnamefont {Shields}},\ }\href@noop {}
  {\bibfield  {journal} {\bibinfo  {journal} {Nature}\ }\textbf {\bibinfo
  {volume} {557}},\ \bibinfo {pages} {400} (\bibinfo {year}
  {2018})}\BibitemShut {NoStop}%
\bibitem [{\citenamefont {Wang}\ \emph {et~al.}(2018)\citenamefont {Wang},
  \citenamefont {Yu},\ and\ \citenamefont {Hu}}]{wang2018twin}%
  \BibitemOpen
  \bibfield  {author} {\bibinfo {author} {\bibfnamefont {X.-B.}\ \bibnamefont
  {Wang}}, \bibinfo {author} {\bibfnamefont {Z.-W.}\ \bibnamefont {Yu}}, \ and\
  \bibinfo {author} {\bibfnamefont {X.-L.}\ \bibnamefont {Hu}},\ }\href@noop {}
  {\bibfield  {journal} {\bibinfo  {journal} {Physical Review A}\ }\textbf
  {\bibinfo {volume} {98}},\ \bibinfo {pages} {062323} (\bibinfo {year}
  {2018})}\BibitemShut {NoStop}%
\bibitem [{\citenamefont {Tamaki}\ \emph {et~al.}(2018)\citenamefont {Tamaki},
  \citenamefont {Lo}, \citenamefont {Wang},\ and\ \citenamefont
  {Lucamarini}}]{tamaki2018information}%
  \BibitemOpen
  \bibfield  {author} {\bibinfo {author} {\bibfnamefont {K.}~\bibnamefont
  {Tamaki}}, \bibinfo {author} {\bibfnamefont {H.-K.}\ \bibnamefont {Lo}},
  \bibinfo {author} {\bibfnamefont {W.}~\bibnamefont {Wang}}, \ and\ \bibinfo
  {author} {\bibfnamefont {M.}~\bibnamefont {Lucamarini}},\ }\href@noop {}
  {\bibfield  {journal} {\bibinfo  {journal} {arXiv preprint arXiv:1805.05511}\
  } (\bibinfo {year} {2018})}\BibitemShut {NoStop}%
\bibitem [{\citenamefont {Ma}\ \emph {et~al.}(2018)\citenamefont {Ma},
  \citenamefont {Zeng},\ and\ \citenamefont {Zhou}}]{ma2018phase}%
  \BibitemOpen
  \bibfield  {author} {\bibinfo {author} {\bibfnamefont {X.}~\bibnamefont
  {Ma}}, \bibinfo {author} {\bibfnamefont {P.}~\bibnamefont {Zeng}}, \ and\
  \bibinfo {author} {\bibfnamefont {H.}~\bibnamefont {Zhou}},\ }\href@noop {}
  {\bibfield  {journal} {\bibinfo  {journal} {Physical Review X}\ }\textbf
  {\bibinfo {volume} {8}},\ \bibinfo {pages} {031043} (\bibinfo {year}
  {2018})}\BibitemShut {NoStop}%
\bibitem [{\citenamefont {Cui}\ \emph {et~al.}(2019)\citenamefont {Cui},
  \citenamefont {Yin}, \citenamefont {Wang}, \citenamefont {Chen},
  \citenamefont {Wang}, \citenamefont {Guo},\ and\ \citenamefont
  {Han}}]{cui2019twin}%
  \BibitemOpen
  \bibfield  {author} {\bibinfo {author} {\bibfnamefont {C.}~\bibnamefont
  {Cui}}, \bibinfo {author} {\bibfnamefont {Z.-Q.}\ \bibnamefont {Yin}},
  \bibinfo {author} {\bibfnamefont {R.}~\bibnamefont {Wang}}, \bibinfo {author}
  {\bibfnamefont {W.}~\bibnamefont {Chen}}, \bibinfo {author} {\bibfnamefont
  {S.}~\bibnamefont {Wang}}, \bibinfo {author} {\bibfnamefont {G.-C.}\
  \bibnamefont {Guo}}, \ and\ \bibinfo {author} {\bibfnamefont {Z.-F.}\
  \bibnamefont {Han}},\ }\href@noop {} {\bibfield  {journal} {\bibinfo
  {journal} {Physical Review Applied}\ }\textbf {\bibinfo {volume} {11}},\
  \bibinfo {pages} {034053} (\bibinfo {year} {2019})}\BibitemShut {NoStop}%
\bibitem [{\citenamefont {Curty}\ \emph {et~al.}(2019)\citenamefont {Curty},
  \citenamefont {Azuma},\ and\ \citenamefont {Lo}}]{curty2019simple}%
  \BibitemOpen
  \bibfield  {author} {\bibinfo {author} {\bibfnamefont {M.}~\bibnamefont
  {Curty}}, \bibinfo {author} {\bibfnamefont {K.}~\bibnamefont {Azuma}}, \ and\
  \bibinfo {author} {\bibfnamefont {H.-K.}\ \bibnamefont {Lo}},\ }\href@noop {}
  {\bibfield  {journal} {\bibinfo  {journal} {npj Quantum Information}\
  }\textbf {\bibinfo {volume} {5}},\ \bibinfo {pages} {1} (\bibinfo {year}
  {2019})}\BibitemShut {NoStop}%
\bibitem [{\citenamefont {Lu}\ \emph {et~al.}(2019)\citenamefont {Lu},
  \citenamefont {Yin}, \citenamefont {Cui}, \citenamefont {Fan-Yuan},
  \citenamefont {Wang}, \citenamefont {Wang}, \citenamefont {Chen},
  \citenamefont {He}, \citenamefont {Huang}, \citenamefont {Xu} \emph
  {et~al.}}]{lu2019improving}%
  \BibitemOpen
  \bibfield  {author} {\bibinfo {author} {\bibfnamefont {F.-Y.}\ \bibnamefont
  {Lu}}, \bibinfo {author} {\bibfnamefont {Z.-Q.}\ \bibnamefont {Yin}},
  \bibinfo {author} {\bibfnamefont {C.-H.}\ \bibnamefont {Cui}}, \bibinfo
  {author} {\bibfnamefont {G.-J.}\ \bibnamefont {Fan-Yuan}}, \bibinfo {author}
  {\bibfnamefont {R.}~\bibnamefont {Wang}}, \bibinfo {author} {\bibfnamefont
  {S.}~\bibnamefont {Wang}}, \bibinfo {author} {\bibfnamefont {W.}~\bibnamefont
  {Chen}}, \bibinfo {author} {\bibfnamefont {D.-Y.}\ \bibnamefont {He}},
  \bibinfo {author} {\bibfnamefont {W.}~\bibnamefont {Huang}}, \bibinfo
  {author} {\bibfnamefont {B.-J.}\ \bibnamefont {Xu}},  \emph {et~al.},\
  }\href@noop {} {\bibfield  {journal} {\bibinfo  {journal} {Physical Review
  A}\ }\textbf {\bibinfo {volume} {100}},\ \bibinfo {pages} {022306} (\bibinfo
  {year} {2019})}\BibitemShut {NoStop}%
\bibitem [{\citenamefont {Maeda}\ \emph {et~al.}(2019)\citenamefont {Maeda},
  \citenamefont {Sasaki},\ and\ \citenamefont
  {Koashi}}]{maeda2019repeaterless}%
  \BibitemOpen
  \bibfield  {author} {\bibinfo {author} {\bibfnamefont {K.}~\bibnamefont
  {Maeda}}, \bibinfo {author} {\bibfnamefont {T.}~\bibnamefont {Sasaki}}, \
  and\ \bibinfo {author} {\bibfnamefont {M.}~\bibnamefont {Koashi}},\
  }\href@noop {} {\bibfield  {journal} {\bibinfo  {journal} {Nature
  communications}\ }\textbf {\bibinfo {volume} {10}},\ \bibinfo {pages} {3140}
  (\bibinfo {year} {2019})}\BibitemShut {NoStop}%
\bibitem [{\citenamefont {Curr{\'a}s-Lorenzo}\ \emph
  {et~al.}(2021)\citenamefont {Curr{\'a}s-Lorenzo}, \citenamefont {Navarrete},
  \citenamefont {Azuma}, \citenamefont {Kato}, \citenamefont {Curty},\ and\
  \citenamefont {Razavi}}]{curras2021tight}%
  \BibitemOpen
  \bibfield  {author} {\bibinfo {author} {\bibfnamefont {G.}~\bibnamefont
  {Curr{\'a}s-Lorenzo}}, \bibinfo {author} {\bibfnamefont {{\'A}.}~\bibnamefont
  {Navarrete}}, \bibinfo {author} {\bibfnamefont {K.}~\bibnamefont {Azuma}},
  \bibinfo {author} {\bibfnamefont {G.}~\bibnamefont {Kato}}, \bibinfo {author}
  {\bibfnamefont {M.}~\bibnamefont {Curty}}, \ and\ \bibinfo {author}
  {\bibfnamefont {M.}~\bibnamefont {Razavi}},\ }\href@noop {} {\bibfield
  {journal} {\bibinfo  {journal} {npj Quantum Information}\ }\textbf {\bibinfo
  {volume} {7}},\ \bibinfo {pages} {1} (\bibinfo {year} {2021})}\BibitemShut
  {NoStop}%
\bibitem [{\citenamefont {Minder}\ \emph {et~al.}(2019)\citenamefont {Minder},
  \citenamefont {Pittaluga}, \citenamefont {Roberts}, \citenamefont
  {Lucamarini}, \citenamefont {Dynes}, \citenamefont {Yuan},\ and\
  \citenamefont {Shields}}]{minder2019experimental}%
  \BibitemOpen
  \bibfield  {author} {\bibinfo {author} {\bibfnamefont {M.}~\bibnamefont
  {Minder}}, \bibinfo {author} {\bibfnamefont {M.}~\bibnamefont {Pittaluga}},
  \bibinfo {author} {\bibfnamefont {G.}~\bibnamefont {Roberts}}, \bibinfo
  {author} {\bibfnamefont {M.}~\bibnamefont {Lucamarini}}, \bibinfo {author}
  {\bibfnamefont {J.}~\bibnamefont {Dynes}}, \bibinfo {author} {\bibfnamefont
  {Z.}~\bibnamefont {Yuan}}, \ and\ \bibinfo {author} {\bibfnamefont
  {A.}~\bibnamefont {Shields}},\ }\href@noop {} {\bibfield  {journal} {\bibinfo
   {journal} {Nature Photonics}\ }\textbf {\bibinfo {volume} {13}},\ \bibinfo
  {pages} {334} (\bibinfo {year} {2019})}\BibitemShut {NoStop}%
\bibitem [{\citenamefont {Liu}\ \emph {et~al.}(2019)\citenamefont {Liu},
  \citenamefont {Yu}, \citenamefont {Zhang}, \citenamefont {Guan},
  \citenamefont {Chen}, \citenamefont {Zhang}, \citenamefont {Hu},
  \citenamefont {Li}, \citenamefont {Jiang}, \citenamefont {Lin} \emph
  {et~al.}}]{liu2019experimental}%
  \BibitemOpen
  \bibfield  {author} {\bibinfo {author} {\bibfnamefont {Y.}~\bibnamefont
  {Liu}}, \bibinfo {author} {\bibfnamefont {Z.-W.}\ \bibnamefont {Yu}},
  \bibinfo {author} {\bibfnamefont {W.}~\bibnamefont {Zhang}}, \bibinfo
  {author} {\bibfnamefont {J.-Y.}\ \bibnamefont {Guan}}, \bibinfo {author}
  {\bibfnamefont {J.-P.}\ \bibnamefont {Chen}}, \bibinfo {author}
  {\bibfnamefont {C.}~\bibnamefont {Zhang}}, \bibinfo {author} {\bibfnamefont
  {X.-L.}\ \bibnamefont {Hu}}, \bibinfo {author} {\bibfnamefont
  {H.}~\bibnamefont {Li}}, \bibinfo {author} {\bibfnamefont {C.}~\bibnamefont
  {Jiang}}, \bibinfo {author} {\bibfnamefont {J.}~\bibnamefont {Lin}},  \emph
  {et~al.},\ }\href@noop {} {\bibfield  {journal} {\bibinfo  {journal}
  {Physical Review Letters}\ }\textbf {\bibinfo {volume} {123}},\ \bibinfo
  {pages} {100505} (\bibinfo {year} {2019})}\BibitemShut {NoStop}%
\bibitem [{\citenamefont {Wang}\ \emph {et~al.}(2019)\citenamefont {Wang},
  \citenamefont {He}, \citenamefont {Yin}, \citenamefont {Lu}, \citenamefont
  {Cui}, \citenamefont {Chen}, \citenamefont {Zhou}, \citenamefont {Guo},\ and\
  \citenamefont {Han}}]{wang2019beating}%
  \BibitemOpen
  \bibfield  {author} {\bibinfo {author} {\bibfnamefont {S.}~\bibnamefont
  {Wang}}, \bibinfo {author} {\bibfnamefont {D.-Y.}\ \bibnamefont {He}},
  \bibinfo {author} {\bibfnamefont {Z.-Q.}\ \bibnamefont {Yin}}, \bibinfo
  {author} {\bibfnamefont {F.-Y.}\ \bibnamefont {Lu}}, \bibinfo {author}
  {\bibfnamefont {C.-H.}\ \bibnamefont {Cui}}, \bibinfo {author} {\bibfnamefont
  {W.}~\bibnamefont {Chen}}, \bibinfo {author} {\bibfnamefont {Z.}~\bibnamefont
  {Zhou}}, \bibinfo {author} {\bibfnamefont {G.-C.}\ \bibnamefont {Guo}}, \
  and\ \bibinfo {author} {\bibfnamefont {Z.-F.}\ \bibnamefont {Han}},\
  }\href@noop {} {\bibfield  {journal} {\bibinfo  {journal} {Physical Review
  X}\ }\textbf {\bibinfo {volume} {9}},\ \bibinfo {pages} {021046} (\bibinfo
  {year} {2019})}\BibitemShut {NoStop}%
\bibitem [{\citenamefont {Zhong}\ \emph {et~al.}(2019)\citenamefont {Zhong},
  \citenamefont {Hu}, \citenamefont {Curty}, \citenamefont {Qian},\ and\
  \citenamefont {Lo}}]{zhong2019proof}%
  \BibitemOpen
  \bibfield  {author} {\bibinfo {author} {\bibfnamefont {X.}~\bibnamefont
  {Zhong}}, \bibinfo {author} {\bibfnamefont {J.}~\bibnamefont {Hu}}, \bibinfo
  {author} {\bibfnamefont {M.}~\bibnamefont {Curty}}, \bibinfo {author}
  {\bibfnamefont {L.}~\bibnamefont {Qian}}, \ and\ \bibinfo {author}
  {\bibfnamefont {H.-K.}\ \bibnamefont {Lo}},\ }\href@noop {} {\bibfield
  {journal} {\bibinfo  {journal} {Physical Review Letters}\ }\textbf {\bibinfo
  {volume} {123}},\ \bibinfo {pages} {100506} (\bibinfo {year}
  {2019})}\BibitemShut {NoStop}%
\bibitem [{\citenamefont {Chen}\ \emph {et~al.}(2020)\citenamefont {Chen},
  \citenamefont {Zhang}, \citenamefont {Liu}, \citenamefont {Jiang},
  \citenamefont {Zhang}, \citenamefont {Hu}, \citenamefont {Guan},
  \citenamefont {Yu}, \citenamefont {Xu}, \citenamefont {Lin}, \citenamefont
  {Li}, \citenamefont {Chen}, \citenamefont {Li}, \citenamefont {You},
  \citenamefont {Wang}, \citenamefont {Wang}, \citenamefont {Zhang},\ and\
  \citenamefont {Pan}}]{chen2020sending}%
  \BibitemOpen
  \bibfield  {author} {\bibinfo {author} {\bibfnamefont {J.-P.}\ \bibnamefont
  {Chen}}, \bibinfo {author} {\bibfnamefont {C.}~\bibnamefont {Zhang}},
  \bibinfo {author} {\bibfnamefont {Y.}~\bibnamefont {Liu}}, \bibinfo {author}
  {\bibfnamefont {C.}~\bibnamefont {Jiang}}, \bibinfo {author} {\bibfnamefont
  {W.}~\bibnamefont {Zhang}}, \bibinfo {author} {\bibfnamefont {X.-L.}\
  \bibnamefont {Hu}}, \bibinfo {author} {\bibfnamefont {J.-Y.}\ \bibnamefont
  {Guan}}, \bibinfo {author} {\bibfnamefont {Z.-W.}\ \bibnamefont {Yu}},
  \bibinfo {author} {\bibfnamefont {H.}~\bibnamefont {Xu}}, \bibinfo {author}
  {\bibfnamefont {J.}~\bibnamefont {Lin}}, \bibinfo {author} {\bibfnamefont
  {M.-J.}\ \bibnamefont {Li}}, \bibinfo {author} {\bibfnamefont
  {H.}~\bibnamefont {Chen}}, \bibinfo {author} {\bibfnamefont {H.}~\bibnamefont
  {Li}}, \bibinfo {author} {\bibfnamefont {L.}~\bibnamefont {You}}, \bibinfo
  {author} {\bibfnamefont {Z.}~\bibnamefont {Wang}}, \bibinfo {author}
  {\bibfnamefont {X.-B.}\ \bibnamefont {Wang}}, \bibinfo {author}
  {\bibfnamefont {Q.}~\bibnamefont {Zhang}}, \ and\ \bibinfo {author}
  {\bibfnamefont {J.-W.}\ \bibnamefont {Pan}},\ }\href@noop {} {\bibfield
  {journal} {\bibinfo  {journal} {Physical Review Letters}\ }\textbf {\bibinfo
  {volume} {124}},\ \bibinfo {pages} {070501} (\bibinfo {year}
  {2020})}\BibitemShut {NoStop}%
\bibitem [{\citenamefont {Liu}\ \emph {et~al.}(2021)\citenamefont {Liu},
  \citenamefont {Jiang}, \citenamefont {Zhu}, \citenamefont {Zou},
  \citenamefont {Yu}, \citenamefont {Hu}, \citenamefont {Xu}, \citenamefont
  {Ma}, \citenamefont {Han}, \citenamefont {Chen} \emph
  {et~al.}}]{liu2021field}%
  \BibitemOpen
  \bibfield  {author} {\bibinfo {author} {\bibfnamefont {H.}~\bibnamefont
  {Liu}}, \bibinfo {author} {\bibfnamefont {C.}~\bibnamefont {Jiang}}, \bibinfo
  {author} {\bibfnamefont {H.-T.}\ \bibnamefont {Zhu}}, \bibinfo {author}
  {\bibfnamefont {M.}~\bibnamefont {Zou}}, \bibinfo {author} {\bibfnamefont
  {Z.-W.}\ \bibnamefont {Yu}}, \bibinfo {author} {\bibfnamefont {X.-L.}\
  \bibnamefont {Hu}}, \bibinfo {author} {\bibfnamefont {H.}~\bibnamefont {Xu}},
  \bibinfo {author} {\bibfnamefont {S.}~\bibnamefont {Ma}}, \bibinfo {author}
  {\bibfnamefont {Z.}~\bibnamefont {Han}}, \bibinfo {author} {\bibfnamefont
  {J.-P.}\ \bibnamefont {Chen}},  \emph {et~al.},\ }\href@noop {} {\bibfield
  {journal} {\bibinfo  {journal} {Physical Review Letters}\ }\textbf {\bibinfo
  {volume} {126}},\ \bibinfo {pages} {250502} (\bibinfo {year}
  {2021})}\BibitemShut {NoStop}%
\bibitem [{\citenamefont {Chen}\ \emph {et~al.}(2021)\citenamefont {Chen},
  \citenamefont {Zhang}, \citenamefont {Liu}, \citenamefont {Jiang},
  \citenamefont {Zhang}, \citenamefont {Han}, \citenamefont {Ma}, \citenamefont
  {Hu}, \citenamefont {Li}, \citenamefont {Liu} \emph {et~al.}}]{chen2021twin}%
  \BibitemOpen
  \bibfield  {author} {\bibinfo {author} {\bibfnamefont {J.-P.}\ \bibnamefont
  {Chen}}, \bibinfo {author} {\bibfnamefont {C.}~\bibnamefont {Zhang}},
  \bibinfo {author} {\bibfnamefont {Y.}~\bibnamefont {Liu}}, \bibinfo {author}
  {\bibfnamefont {C.}~\bibnamefont {Jiang}}, \bibinfo {author} {\bibfnamefont
  {W.-J.}\ \bibnamefont {Zhang}}, \bibinfo {author} {\bibfnamefont {Z.-Y.}\
  \bibnamefont {Han}}, \bibinfo {author} {\bibfnamefont {S.-Z.}\ \bibnamefont
  {Ma}}, \bibinfo {author} {\bibfnamefont {X.-L.}\ \bibnamefont {Hu}}, \bibinfo
  {author} {\bibfnamefont {Y.-H.}\ \bibnamefont {Li}}, \bibinfo {author}
  {\bibfnamefont {H.}~\bibnamefont {Liu}},  \emph {et~al.},\ }\href@noop {}
  {\bibfield  {journal} {\bibinfo  {journal} {Nature Photonics}\ ,\ \bibinfo
  {pages} {1}} (\bibinfo {year} {2021})}\BibitemShut {NoStop}%
\bibitem [{\citenamefont {Pittaluga}\ \emph {et~al.}(2021)\citenamefont
  {Pittaluga}, \citenamefont {Minder}, \citenamefont {Lucamarini},
  \citenamefont {Sanzaro}, \citenamefont {Woodward}, \citenamefont {Li},
  \citenamefont {Yuan},\ and\ \citenamefont {Shields}}]{pittaluga2021600}%
  \BibitemOpen
  \bibfield  {author} {\bibinfo {author} {\bibfnamefont {M.}~\bibnamefont
  {Pittaluga}}, \bibinfo {author} {\bibfnamefont {M.}~\bibnamefont {Minder}},
  \bibinfo {author} {\bibfnamefont {M.}~\bibnamefont {Lucamarini}}, \bibinfo
  {author} {\bibfnamefont {M.}~\bibnamefont {Sanzaro}}, \bibinfo {author}
  {\bibfnamefont {R.~I.}\ \bibnamefont {Woodward}}, \bibinfo {author}
  {\bibfnamefont {M.-J.}\ \bibnamefont {Li}}, \bibinfo {author} {\bibfnamefont
  {Z.}~\bibnamefont {Yuan}}, \ and\ \bibinfo {author} {\bibfnamefont {A.~J.}\
  \bibnamefont {Shields}},\ }\href@noop {} {\bibfield  {journal} {\bibinfo
  {journal} {Nature Photonics}\ }\textbf {\bibinfo {volume} {15}},\ \bibinfo
  {pages} {530} (\bibinfo {year} {2021})}\BibitemShut {NoStop}%
\bibitem [{\citenamefont {Wang}\ \emph {et~al.}(2022)\citenamefont {Wang},
  \citenamefont {Yin}, \citenamefont {He}, \citenamefont {Chen}, \citenamefont
  {Wang}, \citenamefont {Ye}, \citenamefont {Zhou}, \citenamefont {Fan-Yuan},
  \citenamefont {Wang}, \citenamefont {Chen}, \citenamefont {Zhu},
  \citenamefont {Morozov}, \citenamefont {Divochiy}, \citenamefont {Zhou},
  \citenamefont {Guo},\ and\ \citenamefont {Han}}]{wang2022twin}%
  \BibitemOpen
  \bibfield  {author} {\bibinfo {author} {\bibfnamefont {S.}~\bibnamefont
  {Wang}}, \bibinfo {author} {\bibfnamefont {Z.-Q.}\ \bibnamefont {Yin}},
  \bibinfo {author} {\bibfnamefont {D.-Y.}\ \bibnamefont {He}}, \bibinfo
  {author} {\bibfnamefont {W.}~\bibnamefont {Chen}}, \bibinfo {author}
  {\bibfnamefont {R.-Q.}\ \bibnamefont {Wang}}, \bibinfo {author}
  {\bibfnamefont {P.}~\bibnamefont {Ye}}, \bibinfo {author} {\bibfnamefont
  {Y.}~\bibnamefont {Zhou}}, \bibinfo {author} {\bibfnamefont {G.-J.}\
  \bibnamefont {Fan-Yuan}}, \bibinfo {author} {\bibfnamefont {F.-X.}\
  \bibnamefont {Wang}}, \bibinfo {author} {\bibfnamefont {W.}~\bibnamefont
  {Chen}}, \bibinfo {author} {\bibfnamefont {Y.-G.}\ \bibnamefont {Zhu}},
  \bibinfo {author} {\bibfnamefont {P.~V.}\ \bibnamefont {Morozov}}, \bibinfo
  {author} {\bibfnamefont {A.~V.}\ \bibnamefont {Divochiy}}, \bibinfo {author}
  {\bibfnamefont {Z.}~\bibnamefont {Zhou}}, \bibinfo {author} {\bibfnamefont
  {G.-C.}\ \bibnamefont {Guo}}, \ and\ \bibinfo {author} {\bibfnamefont
  {Z.-F.}\ \bibnamefont {Han}},\ }\href@noop {} {\bibfield  {journal} {\bibinfo
   {journal} {Nature Photonics, published online}\ } (\bibinfo {year}
  {2022})}\BibitemShut {NoStop}%
\bibitem [{\citenamefont {Clivati}\ \emph {et~al.}(2022)\citenamefont
  {Clivati}, \citenamefont {Meda}, \citenamefont {Donadello}, \citenamefont
  {Virz{\`\i}}, \citenamefont {Genovese}, \citenamefont {Levi}, \citenamefont
  {Mura}, \citenamefont {Pittaluga}, \citenamefont {Yuan}, \citenamefont
  {Shields}, \citenamefont {Lucamarini}, \citenamefont {Degiovanni},\ and\
  \citenamefont {Calonico}}]{clivati2022coherent}%
  \BibitemOpen
  \bibfield  {author} {\bibinfo {author} {\bibfnamefont {C.}~\bibnamefont
  {Clivati}}, \bibinfo {author} {\bibfnamefont {A.}~\bibnamefont {Meda}},
  \bibinfo {author} {\bibfnamefont {S.}~\bibnamefont {Donadello}}, \bibinfo
  {author} {\bibfnamefont {S.}~\bibnamefont {Virz{\`\i}}}, \bibinfo {author}
  {\bibfnamefont {M.}~\bibnamefont {Genovese}}, \bibinfo {author}
  {\bibfnamefont {F.}~\bibnamefont {Levi}}, \bibinfo {author} {\bibfnamefont
  {A.}~\bibnamefont {Mura}}, \bibinfo {author} {\bibfnamefont {M.}~\bibnamefont
  {Pittaluga}}, \bibinfo {author} {\bibfnamefont {Z.}~\bibnamefont {Yuan}},
  \bibinfo {author} {\bibfnamefont {A.~J.}\ \bibnamefont {Shields}}, \bibinfo
  {author} {\bibfnamefont {M.}~\bibnamefont {Lucamarini}}, \bibinfo {author}
  {\bibfnamefont {I.~P.}\ \bibnamefont {Degiovanni}}, \ and\ \bibinfo {author}
  {\bibfnamefont {D.}~\bibnamefont {Calonico}},\ }\href@noop {} {\bibfield
  {journal} {\bibinfo  {journal} {Nature Communications}\ }\textbf {\bibinfo
  {volume} {13}},\ \bibinfo {pages} {1} (\bibinfo {year} {2022})}\BibitemShut
  {NoStop}%
\bibitem [{\citenamefont {Yu}\ \emph {et~al.}(2019)\citenamefont {Yu},
  \citenamefont {Hu}, \citenamefont {Jiang}, \citenamefont {Xu},\ and\
  \citenamefont {Wang}}]{yu2019sending}%
  \BibitemOpen
  \bibfield  {author} {\bibinfo {author} {\bibfnamefont {Z.-W.}\ \bibnamefont
  {Yu}}, \bibinfo {author} {\bibfnamefont {X.-L.}\ \bibnamefont {Hu}}, \bibinfo
  {author} {\bibfnamefont {C.}~\bibnamefont {Jiang}}, \bibinfo {author}
  {\bibfnamefont {H.}~\bibnamefont {Xu}}, \ and\ \bibinfo {author}
  {\bibfnamefont {X.-B.}\ \bibnamefont {Wang}},\ }\href@noop {} {\bibfield
  {journal} {\bibinfo  {journal} {Scientific reports}\ }\textbf {\bibinfo
  {volume} {9}},\ \bibinfo {pages} {3080} (\bibinfo {year} {2019})}\BibitemShut
  {NoStop}%
\bibitem [{\citenamefont {Jiang}\ \emph {et~al.}(2019)\citenamefont {Jiang},
  \citenamefont {Yu}, \citenamefont {Hu},\ and\ \citenamefont
  {Wang}}]{jiang2019unconditional}%
  \BibitemOpen
  \bibfield  {author} {\bibinfo {author} {\bibfnamefont {C.}~\bibnamefont
  {Jiang}}, \bibinfo {author} {\bibfnamefont {Z.-W.}\ \bibnamefont {Yu}},
  \bibinfo {author} {\bibfnamefont {X.-L.}\ \bibnamefont {Hu}}, \ and\ \bibinfo
  {author} {\bibfnamefont {X.-B.}\ \bibnamefont {Wang}},\ }\href@noop {}
  {\bibfield  {journal} {\bibinfo  {journal} {Physical Review Applied}\
  }\textbf {\bibinfo {volume} {12}},\ \bibinfo {pages} {024061} (\bibinfo
  {year} {2019})}\BibitemShut {NoStop}%
\bibitem [{\citenamefont {Hu}\ \emph {et~al.}(2019)\citenamefont {Hu},
  \citenamefont {Jiang}, \citenamefont {Yu},\ and\ \citenamefont
  {Wang}}]{hu2019sending}%
  \BibitemOpen
  \bibfield  {author} {\bibinfo {author} {\bibfnamefont {X.-L.}\ \bibnamefont
  {Hu}}, \bibinfo {author} {\bibfnamefont {C.}~\bibnamefont {Jiang}}, \bibinfo
  {author} {\bibfnamefont {Z.-W.}\ \bibnamefont {Yu}}, \ and\ \bibinfo {author}
  {\bibfnamefont {X.-B.}\ \bibnamefont {Wang}},\ }\href@noop {} {\bibfield
  {journal} {\bibinfo  {journal} {Physical Review A}\ }\textbf {\bibinfo
  {volume} {100}},\ \bibinfo {pages} {062337} (\bibinfo {year}
  {2019})}\BibitemShut {NoStop}%
\bibitem [{\citenamefont {Xu}\ \emph {et~al.}(2020{\natexlab{b}})\citenamefont
  {Xu}, \citenamefont {Yu}, \citenamefont {Jiang}, \citenamefont {Hu},\ and\
  \citenamefont {Wang}}]{xu2020sending}%
  \BibitemOpen
  \bibfield  {author} {\bibinfo {author} {\bibfnamefont {H.}~\bibnamefont
  {Xu}}, \bibinfo {author} {\bibfnamefont {Z.-W.}\ \bibnamefont {Yu}}, \bibinfo
  {author} {\bibfnamefont {C.}~\bibnamefont {Jiang}}, \bibinfo {author}
  {\bibfnamefont {X.-L.}\ \bibnamefont {Hu}}, \ and\ \bibinfo {author}
  {\bibfnamefont {X.-B.}\ \bibnamefont {Wang}},\ }\href@noop {} {\bibfield
  {journal} {\bibinfo  {journal} {Physical Review A}\ }\textbf {\bibinfo
  {volume} {101}},\ \bibinfo {pages} {042330} (\bibinfo {year}
  {2020}{\natexlab{b}})}\BibitemShut {NoStop}%
\bibitem [{\citenamefont {Jiang}\ \emph {et~al.}(2020)\citenamefont {Jiang},
  \citenamefont {Hu}, \citenamefont {Xu}, \citenamefont {Yu},\ and\
  \citenamefont {Wang}}]{jiang2020zigzag}%
  \BibitemOpen
  \bibfield  {author} {\bibinfo {author} {\bibfnamefont {C.}~\bibnamefont
  {Jiang}}, \bibinfo {author} {\bibfnamefont {X.-L.}\ \bibnamefont {Hu}},
  \bibinfo {author} {\bibfnamefont {H.}~\bibnamefont {Xu}}, \bibinfo {author}
  {\bibfnamefont {Z.-W.}\ \bibnamefont {Yu}}, \ and\ \bibinfo {author}
  {\bibfnamefont {X.-B.}\ \bibnamefont {Wang}},\ }\href@noop {} {\bibfield
  {journal} {\bibinfo  {journal} {New Journal of Physics}\ }\textbf {\bibinfo
  {volume} {22}},\ \bibinfo {pages} {053048} (\bibinfo {year}
  {2020})}\BibitemShut {NoStop}%
\bibitem [{\citenamefont {Jiang}\ \emph
  {et~al.}(2021{\natexlab{b}})\citenamefont {Jiang}, \citenamefont {Hu},
  \citenamefont {Yu},\ and\ \citenamefont {Wang}}]{jiang2021composable}%
  \BibitemOpen
  \bibfield  {author} {\bibinfo {author} {\bibfnamefont {C.}~\bibnamefont
  {Jiang}}, \bibinfo {author} {\bibfnamefont {X.-L.}\ \bibnamefont {Hu}},
  \bibinfo {author} {\bibfnamefont {Z.-W.}\ \bibnamefont {Yu}}, \ and\ \bibinfo
  {author} {\bibfnamefont {X.-B.}\ \bibnamefont {Wang}},\ }\href@noop {}
  {\bibfield  {journal} {\bibinfo  {journal} {New Journal of Physics}\ }\textbf
  {\bibinfo {volume} {23}},\ \bibinfo {pages} {063038} (\bibinfo {year}
  {2021}{\natexlab{b}})}\BibitemShut {NoStop}%
\bibitem [{\citenamefont {Teng}\ \emph {et~al.}(2021)\citenamefont {Teng},
  \citenamefont {Yin}, \citenamefont {Fan-Yuan}, \citenamefont {Lu},
  \citenamefont {Wang}, \citenamefont {Wang}, \citenamefont {Chen},
  \citenamefont {Huang}, \citenamefont {Xu}, \citenamefont {Guo} \emph
  {et~al.}}]{teng2021sending}%
  \BibitemOpen
  \bibfield  {author} {\bibinfo {author} {\bibfnamefont {J.}~\bibnamefont
  {Teng}}, \bibinfo {author} {\bibfnamefont {Z.-Q.}\ \bibnamefont {Yin}},
  \bibinfo {author} {\bibfnamefont {G.-J.}\ \bibnamefont {Fan-Yuan}}, \bibinfo
  {author} {\bibfnamefont {F.-Y.}\ \bibnamefont {Lu}}, \bibinfo {author}
  {\bibfnamefont {R.}~\bibnamefont {Wang}}, \bibinfo {author} {\bibfnamefont
  {S.}~\bibnamefont {Wang}}, \bibinfo {author} {\bibfnamefont {W.}~\bibnamefont
  {Chen}}, \bibinfo {author} {\bibfnamefont {W.}~\bibnamefont {Huang}},
  \bibinfo {author} {\bibfnamefont {B.-J.}\ \bibnamefont {Xu}}, \bibinfo
  {author} {\bibfnamefont {G.-C.}\ \bibnamefont {Guo}},  \emph {et~al.},\
  }\href@noop {} {\bibfield  {journal} {\bibinfo  {journal} {Physical Review
  A}\ }\textbf {\bibinfo {volume} {104}},\ \bibinfo {pages} {062441} (\bibinfo
  {year} {2021})}\BibitemShut {NoStop}%
\bibitem [{\citenamefont {Chernoff}(1952)}]{chernoff1952measure}%
  \BibitemOpen
  \bibfield  {author} {\bibinfo {author} {\bibfnamefont {H.}~\bibnamefont
  {Chernoff}},\ }\href@noop {} {\bibfield  {journal} {\bibinfo  {journal} {The
  Annals of Mathematical Statistics}\ }\textbf {\bibinfo {volume} {23}},\
  \bibinfo {pages} {493} (\bibinfo {year} {1952})}\BibitemShut {NoStop}%
\bibitem [{\citenamefont {Lo}(2005)}]{lo2005getting}%
  \BibitemOpen
  \bibfield  {author} {\bibinfo {author} {\bibfnamefont {H.-K.}\ \bibnamefont
  {Lo}},\ }\href@noop {} {\bibfield  {journal} {\bibinfo  {journal} {Quantum
  Information and Computation}\ }\textbf {\bibinfo {volume} {5}},\ \bibinfo
  {pages} {413} (\bibinfo {year} {2005})}\BibitemShut {NoStop}%
\bibitem [{\citenamefont {Chau}(2020)}]{chau2020security}%
  \BibitemOpen
  \bibfield  {author} {\bibinfo {author} {\bibfnamefont {H.}~\bibnamefont
  {Chau}},\ }\href@noop {} {\bibfield  {journal} {\bibinfo  {journal} {Physical
  Review A}\ }\textbf {\bibinfo {volume} {102}},\ \bibinfo {pages} {012611}
  (\bibinfo {year} {2020})}\BibitemShut {NoStop}%
\end{thebibliography}%

\end{document}